\documentclass{article}

\usepackage{amsmath}
\usepackage{amsthm}
\usepackage{amssymb}
\usepackage{geometry}
\usepackage{graphicx}
\usepackage{longtable}
\usepackage{booktabs}
\usepackage{natbib}
\usepackage{subcaption}
\usepackage[colorlinks]{hyperref}
\usepackage{stmaryrd}
\usepackage{algorithm}
\usepackage[noend]{algpseudocode}
\usepackage{tikz}
\usepackage{forest}
\usepackage{enumitem}
\usepackage{placeins}

\theoremstyle{definition}

\newtheorem{example}{Example}[section]
\newtheorem{definition}{Definition}[section]

\DeclareMathOperator{\IN}{~\textbf{in}~}
\DeclareMathOperator{\AND}{~\textbf{and}~}

\newcommand{\APPEND}[2]{\textbf{append}~#1~\textbf{to}~#2}

\newcommand{\glb}{\textit{rul}}
\newcommand{\llb}{\textit{rlb}}

\newcommand{\plusplus}{\mathrel{\textrm{++}}}
\newcommand{\lbheur}{l^\mathrm{heuristic}}
\newcommand{\lbeq}{l^\mathrm{EQ}}
\newcommand{\lbchild}{l_\mathrm{child}}
\newcommand{\lbelim}{l^\mathrm{elim}}
\newcommand{\lbquota}{l^\mathrm{quota}}
\newcommand{\lbdisp}{l^\mathrm{disp}}

\usepackage{natbib}
 \bibpunct[, ]{(}{)}{,}{a}{}{,}%
\newcommand{\election}{\mathcal{E}}
\newcommand{\cands}{\mathcal{C}}
\newcommand{\winners}{\mathcal{W}}
\newcommand{\ballots}{\mathcal{B}}
\newcommand{\quota}{Q}
\newcommand{\seats}{N}
\newcommand{\order}{\pi}

\newcommand{\tally}{V}
\newcommand{\tallymax}[2]{\tally^{\mathrm{max}}_{#1,#2}}
\newcommand{\tallymin}[2]{\tally^{\mathrm{min}}_{#1,#2}}

\newcommand{\bvalue}{B}

\newcommand{\bvaluemin}[2]{\bvalue^{\mathrm{min}}_{#1,#2}}
\newcommand{\bvaluemax}[2]{\bvalue^{\mathrm{max}}_{#1,#2}}

\newcommand{\DT}{\mathrm{DistanceTo}^R_{STV}}
\newcommand{\stand}{\mathcal{S}}
\newcommand{\bigo}{\mathcal{O}}
\newcommand{\OLD}{BST-19}
\newcommand{\pile}[2]{\mathrm{pile}_{#1,#2}}

\newcommand{\tail}{\mathrm{tail}}

\begin{document}

\title{Efficient Lower Bounding of Single Transferable Vote Election Margins}

\author{%
Michelle Blom%
\thanks{Department of Computing and Information Systems, University of
Melbourne} \\
{\small\url{michelle.blom@unimelb.edu.au}}
\and
Alexander Ek%
\thanks{Department of Econometrics and Business Statistics, Monash University}
\\
{\small\url{alexander.ek@monash.edu}}
\and
Peter J. Stuckey%
\thanks{Department of Data Science and Artificial Intelligence, Monash University}
\\
{\small\url{peter.stuckey@monash.edu}}
\and
Vanessa Teague%
\thanks{Thinking Cybersecurity Pty. Ltd., Melbourne, and the Australian
National University} \\
{\small\url{vanessa.teague@anu.edu.au}}
\and
Damjan Vukcevic%
\thanks{Department of Econometrics and Business Statistics, Monash University}
\\
{\small\url{damjan.vukcevic@monash.edu}}
}

\date{25 March 2025}

\maketitle

\begin{abstract}
The \emph{single transferable vote} (STV) is a system of preferential
proportional voting employed in multi-seat elections.  Each ballot cast by
a voter is a (potentially partial) ranking over a set of candidates.  The
\emph{margin of victory}, or simply \emph{margin}, is the smallest number
of ballots that need to be manipulated to alter the set of winners.
Knowledge of the margin of an election gives greater insight into both how
much time and money should be spent on auditing the election, and whether
uncovered mistakes throw the election result into doubt---requiring a
costly repeat election---or can be safely ignored without compromising the
integrity of the result.  Lower bounds on the margin can also be used for
this purpose, in cases where exact margins are difficult to compute.  There
is one existing approach to computing lower bounds on the margin of STV
elections, while there are multiple approaches to finding upper bounds.  In
this paper, we present improvements to this existing lower bound
computation method for STV margins.  The improvements lead to increased
computational efficiency and, in many cases, to the algorithm computing
tighter (higher) lower bounds.
\end{abstract}

\noindent\textbf{Funding.}
This research was supported by the Australian Research Council (Discovery
Project DP220101012, OPTIMA ITTC IC200100009).
\bigskip


\pagebreak

\tableofcontents

% ---------------------------------------------------------------------------

\section{Introduction}

The \emph{single transferable vote} (STV), also called proportional-ranked
choice voting, is an electoral system (i.e., a family of social choice
functions) where voters rank candidates in order of preference and multiple
candidates are then elected in a manner reflecting voter preferences
proportionally. STV is used world-wide, including in Australia (national,
state, and local level); Ireland and Malta (EU, national, and local level); as
well as New Zealand, Northern Ireland, Scotland, and the United States of
America (local level).  It is also used within legislative bodies to elect
officials to particular positions in India, Ireland, Nepal, and Pakistan.

STV tabulation proceeds in rounds, with each round either seating  or
eliminating a candidate. In each case, ballots sitting in the seated or
eliminated candidate's tally pile are transferred to the next-ranked (eligible)
candidate. When ballots are transferred from an elected candidate's tally, they
are reduced in value according to a \textit{transfer value}. This reflects the
notion that a portion of the ballot has contributed to electing the candidate,
with the `unused' portion then distributed to another candidate.

While STV has many desirable properties from a social choice standpoint, it is
notoriously hard for mathematical analysis.  This includes computing the
\emph{margin}, which is the minimum number of ballots that need to be
altered---by changing the marked preferences on the ballots---to change who
wins.\footnote{%
A more general version of this definition also allows alterations where ballots
can be completely removed or added.  Here we only allow alterations that keep
the total number of ballots fixed.}
Understanding the margin of an election is helpful because it tells us how
close an election was.  For example, an election with a margin of 1,000 ballots
tells us that problems affecting the interpretation of less than 1,000 ballots
could not have changed who won.  Similarly, it can aid post-election auditing
efforts, such as risk-limiting audits, a crucial component of evidence-based
elections~\citep{starkWagner12,appelStark20}.  These are increasingly needed as
elections are challenged in democracies around the world.  A risk-limiting
audit is a process designed to efficiently provide affirmative statistical
evidence that the reported winners really won, and correct the outcome (with a
guaranteed high probability) if they did not win.

\cite{xia2012computing} showed that exact computation of the margin for
instant-runoff voting (IRV) elections, a single-winner form of STV, is NP-hard.
Exact computation of the margin for STV is consequently at least NP-hard.
\citet{blom2019toward} presented a best-first branch-and-bound algorithm,
\OLD{}, for computing the exact margin of an STV election by searching over a
tree of possible tabulation \textit{prefixes}. For each round of tabulation
that has occurred so far, a \textit{prefix} defines who was seated or
eliminated in each of those rounds. This tree captured not only what occurred
in the reported tabulation, but outcomes that \textit{could} occur if the cast
ballots were manipulated.  The algorithm involved several components: two
methods for computing an upper bound on the margin; a mixed-integer non-linear
program (MINLP) for computing a minimal manipulation to the ballots cast in an
STV election to realise a specific complete outcome; and two heuristics for
computing a lower bound on the number of ballots cast that would have to be
altered to realise an outcome that \textit{starts} in a specific sequence of
seatings and eliminations. These lower bounding heuristics were used to both
prune portions of the branch-and-bound search space, and  reduce the number
MINLP solves required throughout the algorithm.  \OLD{} was capable of
computing exact margins only for 2-seat STV elections.  A modification  was
proposed in which the method for computing minimal manipulations was replaced
with a relaxation that computed a \textit{lower bound} on the manipulation
required to realise an alternate outcome.  The output of the algorithm then
became a lower bound on the margin.

There are a few limitations of the \OLD{} method that we address in this paper.
First, its heuristics for computing lower bounds on the manipulation required
to realise a specific outcome, or partial outcome, used unsophisticated
reasoning about the potential transfer values of ballots. These heuristics
relied on computing minimum and maximum bounds on candidate tallies in various
contexts, and assumed that all ballots that may have a fractional value did not
contribute to minimum tallies, but contributed to maximum tallies with a value
of 1. This resulted in loose lower bounds, where stronger bounds would make the
method more efficient by pruning more of the search space.  Second, \OLD{} did
not reason about the potential downstream cost of realising a given prefix
through manipulation. A partial outcome that has not yet changed who wins could
be assigned a lower bound on manipulation of zero ballots, yet we know that
some original winner has to be `displaced' in future in favour of an original
loser by a non-zero manipulation.  Third, \OLD{} did not leverage any
structural equivalence or similarities when considering partial outcomes that
were in effect similar to those that had previously been visited and evaluated.
This led to similar problems being re-solved multiple times, wasting time and
resources.

In this this paper, we revisit the problem of calculating lower bounds on the
margin of STV elections, building upon  \OLD{}.  We present several
improvements addressing the aforementioned limitations, allowing us to find
tighter (higher) lower bounds on STV margins, and to do so more quickly.  We
show that our improvements lead to increased computational efficiency, and in
many cases to the algorithm computing tighter lower bounds.  For small
elections, in conjunction with existing upper bounding approaches, the new
algorithm is more frequently able to compute exact margins of victory. Our main
contributions are:
\begin{itemize}
\item \textit{Improved transfer path reasoning:}
    By reasoning over how ballots could have been transferred in a prefix
    (partial outcome defining a sequence of seatings and eliminations), we
    determine (smaller) maximum and (greater) minimum candidate tallies in
    each round of the prefix. This helps us compute tighter lower bounds on
    the manipulation required to realise outcomes that start with the prefix.
    (\autoref{sec:revisedlowerbounds})
\item \textit{A displacement lower bounding heuristic:}
    For a given prefix, we compute a lower bound on the cost of seating a
    reported (original) loser or eliminating a reported (original) winner in
    any election outcome that \textit{completes} the prefix.
    (\autoref{sec:displacement-lb})
\item \textit{Leveraging structural equivalence and dominance:}
    Before adding new partial or complete outcomes to the algorithm's
    branch-and-bound search tree, we check for structural equivalence of these
    outcomes with previously explored and evaluated ones.  New outcomes that
    are dominated by ones we have already explored are pruned from the search space.
    (\autoref{sec:lse})
\end{itemize}
We additionally make use of a new STV margin upper bounding algorithm from the
literature \citep{blom2020did,teague2022annexure}, developed since the
publication of \OLD{}.

The rest of the paper is structured as follows.
In \autoref{sec:preliminaries} we describe the STV tabulation process, and in
\autoref{sec:prior-work} we explain prior work.
In \autoref{sec:new-algorithm} we present our improvements to the \OLD{}
algorithm, followed by experimental results in \autoref{sec:results} and
conclusions in \autoref{sec:conclusion}.

% ---------------------------------------------------------------------------

\section{Preliminaries}\label{sec:preliminaries}

We describe STV and STV tabulation, alongside  mathematical notation that will be used throughout this paper.
We clarify which version of STV we consider, as there are many variations in use world-wide.

\subsection{Single Transferable Vote}\label{sec:stv}

STV is a multi-winner ranked choice (preferential) and proportional election system.
Voters rank candidates in order of preference, from first to last, in either a total order or leaving some candidates unranked, depending on the jurisdiction.
A key complexity of STV is that cast ballots change in value throughout  tabulation.
Each ballot starts with a value of 1, which is subsequently reduced if the ballot is used to elect a candidate to a seat.
To be seated, a candidate's tally must reach or exceed a predefined threshold, known as
the \textit{quota} (also called the \textit{election threshold}).
The \textit{Droop quota}
is typically used, usually defined as follows:\footnote{%
There are a few variations used around the world that differ in terms of
rounding and the use of the `$+1$' terms.  We are using the definition most
commonly found in practice, including in Scotland's council-level
elections (The Scottish Local Government Elections Order 2007, No.\ 42,
Schedule 1, Part III, \S 46) and Australia's federal elections (Commonwealth
Electoral Act 1918, Compilation No.\ 77, Part XVIII, \S 273(8)).}
\begin{equation}
    \text{quota} \ = \ \left\lfloor \frac{\text{\# of validly cast ballots}}{\text{\# of seats} + 1} \right\rfloor + 1.
    \label{eqn:Droop}
\end{equation}

Throughout tabulation, each candidate has a \emph{pile} of ballots, with each ballot  associated with a \emph{ballot value} and each candidate's \emph{tally}  defined as the sum of the ballot values in their pile.
Initially, each candidate is awarded all ballots on which they are ranked \textit{first}, forming their \textit{first preference tally}.
Ballots that fail to rank any candidates (i.e., blank votes), have
uninterpretable first preference votes (e.g., ranks multiple candidates as
first preference), or otherwise fail to follow the rules of the jurisdiction in
question, are discarded.

Tabulation proceeds in rounds in which a single candidate is seated or eliminated, until all seats are filled.
Seating and elimination of a candidate results in the ballots in their pile being moved to other candidates' piles or discarded, as detailed in~\autoref{fig:tabulation} of \autoref{sec:STValg}.
If the number of unfilled seats at some point equals the number of remaining candidates, we seat all remaining candidates.
If no candidate has a quota, we eliminate the candidate with the lowest tally (breaking ties as defined by the jurisdiction in question), moving all ballots in their pile to the next most preferred remaining candidate on the ballot who is eligible to receive those ballots. If no such candidate exists, the ballot \textit{exhausts} or is \textit{exhausted}.
A \emph{remaining} candidate is defined as one that has yet to either be seated or eliminated. An \textit{eligible}  candidate is a remaining candidate  that does not already have a quota's worth of votes at the start of the round.\footnote{%
In some jurisdictions, candidates can become ineligible mid-transfer if they reach a quota mid-transfer.
For simplicity and mathematical convenience, we assume all ballots are transferred instantaneously in a single round, avoiding any mid-transfer ineligibility.
}
Where multiple candidates achieve a quota simultaneously, they are seated in order of their tally, highest to smallest.

When seating a candidate, we use the following process to determine what to do
with the ballots in their pile.
If a candidate receives exactly the number of votes required to be seated,
their ballots  are reduced to value 0 and become exhausted.
However, if the candidate received more votes than needed (a tally greater than
the quota), then the ballots continue in the tabulation, now reduced to be
essentially worth their `unused' portion.
This is determined by the \emph{transfer value}, defined as follows:
\begin{equation}
    \text{transfer value} \ = \ \frac{\text{tally} - \text{quota}}{\text{tally}}.
    \label{eqn:TransferValue_simple}
\end{equation}
Each of these ballots is transferred to the next most preferred eligible
candidate on the ballot. Its new value is equal to its current value multiplied
by the transfer value.
For eliminations, the ballots are transferred with their current ballot value.
This variant of STV is known as the Weighted Inclusive Gregory method.\footnote{%
There are many varying approaches for how to define transfer values, even
across different jurisdictions in the same country.
In this paper we use the \emph{Weighted Inclusive Gregory method}, which is
used in the USA and Scotland and is mathematically convenient to work with.
In Australia, the \emph{Unweighted Inclusive Gregory method} is generally, but
not ubiquitously, used, in which a seated candidates surplus is divided by the
total number of ballots in their tally pile.}

% Total ballots: 1230
\begin{table}[t]
\caption{3-seat STV election between candidates \texttt{A}--\texttt{E}, quota
of 308 votes, stating the (a)~count of ballots with each listed ranking,
and (b)~tallies after each round of counting, noting when quotas were
reached (in bold).}
\label{tab:EGSTV1}
    \begin{subtable}[t]{.3\columnwidth}
      \caption{}
      \label{tab:EGSTV1a}
      \centering
        \begin{tabular}{lr}
\toprule
Ranking & Count \\
\midrule
{}[\texttt{A}]       & 250 \\
{}[\texttt{B, A, C}] & 120 \\
{}[\texttt{C, D}]    & 400 \\
{}[\texttt{E}]       & 350 \\
{}[\texttt{C, E, D}] & 110 \\
\bottomrule
\end{tabular}
    \end{subtable}
    \begin{subtable}[t]{.7\columnwidth}
      \caption{}
      \label{tab:EGSTV1b}
      \centering
        \begin{tabular}{crrrr}
%$\seats$: 3 & $\quota$: 308  & & &\\
\toprule
Candidate & Round 1 & Round 2 & Round 3 & Round 4\\
\midrule
 & \texttt{C} elected    & \texttt{E} elected
 & \texttt{B} eliminated & \texttt{A} elected \\
 & $\tau_1 = 0.396$ & $\tau_2 = 0.12$  &  & \\
\midrule
\texttt{A} &         250  &         250    & 250    & \textbf{370}   \\
\texttt{B} &         120  &         120    & 120    &         ---    \\
\texttt{C} & \textbf{510} &         ---    & ---    &         ---    \\
\texttt{D} &           0  &         201.96 & 201.96 &         201.96 \\
\texttt{E} & \textbf{350} & \textbf{350}   & ---    &         ---    \\
\bottomrule
\end{tabular}
    \end{subtable}
\end{table}

\begin{example}
Consider the 3-seat STV election between candidates \texttt{A} to \texttt{E} in \autoref{tab:EGSTV1}, with 1230 validly cast ballots and a  quota of 308 votes.
The first preference tallies of \texttt{A} to \texttt{E} are 250, 120, 510, 0
    and 350 votes, respectively. Candidates \texttt{C} and \texttt{E} have a
    quota's worth of votes on first preferences.
Candidate \texttt{C} has the largest surplus, at 202 votes, and is elected first.
Their transfer value is $\tau_1 = 202/510 = 0.396$.
The 400 [\texttt{C, D}] ballots are each given a weight of 0.396, and a total of 158.4 votes are added to \texttt{D}'s tally.
The 110 [\texttt{C, E, D}] ballots are each given a weight of 0.396, and are also given to candidate \texttt{D}, skipping \texttt{E} as they already have a quota.
Candidate \texttt{D} now has a tally of 201.96 votes.
Candidate \texttt{E} is then elected. Their transfer value would be $\tau_2 = 42/350 = 0.12$, but all of the ballots in their tally exhaust.
In the third round, no candidate has a quota's worth of votes, so the candidate
    with the smallest tally, \texttt{B}, is eliminated.
The 120 [\texttt{B, A, C}] ballots go to \texttt{A}, each retaining their current value of 1.
At the start of the fourth round, candidate \texttt{A} has reached a quota, at 370 votes, and is elected to the third and final seat. 
\end{example}

\subsection{Mathematical Notation}\label{sec:notation}

We reuse mathematical notation for STV from 
\citet{blom2019toward}, with some minor modifications.

\begin{definition}[STV Election]
An STV election is defined as a tuple $\election = (\cands, \ballots, \seats, \quota, \winners)$ where $\cands$ is the set of candidates up for election, $\ballots$ the multi-set of ballots cast in the election, $\seats$ the number of seats to be filled, $\quota$ the election quota (\autoref{eqn:Droop}), and $\winners$ the subset of candidates elected to a seat (the winners). 
Each ballot $b \in \ballots$ is a partial or complete ranking over the candidates $\cands$.
\end{definition}

\begin{definition}[Margin]
The margin of victory for an STV election $\election = (\cands, \ballots, \seats, \quota, \winners)$ is defined as the smallest number of ballot manipulations required to ensure that a set of candidates $\winners'  \neq \winners$ is elected to a seat (i.e., at least one candidate in $\winners'$ must not appear in $\winners$). A single manipulation changes the ranking on a single ballot $b \in \ballots$ to an alternate ranking. For example, consider a ballot with ranking [\texttt{A}, \texttt{B}, \texttt{C}]. Replacing $b$’s ranking with [\texttt{D}, \texttt{A}] represents a single manipulation.
\end{definition}

\begin{definition}[Election order]
Given an STV election $\election = (\cands, \ballots, \seats, \quota, \winners)$, we represent the outcome of $\election$ as an \emph{election order} $\order$, where $\order$ is a sequence of tuples ($c$, $a$) with $c \in \cands$ and $a \in \{0,1\}.$ The tuple ($c$, 1) denotes that candidate $c$ is elected to a seat, while ($c$, 0) that $c$ has been eliminated. The order $\order$ $=$ [(\texttt{A}, 0), (\texttt{C}, 1), (\texttt{B}, 0), (\texttt{D}, 1)] indicates that candidate \texttt{A} is eliminated in the first round of counting, \texttt{C} is next elected to a seat, \texttt{B} is then eliminated, and then \texttt{D} is elected to a seat.
An order $\order$ is \textit{complete} if it involves the election of $\seats$ candidates, and \textit{partial} if fewer than $\seats$ candidates have been elected in $\order$. 
\end{definition}

An election $\election = (\cands, \ballots, \seats, \quota, \winners)$ is tabulated in rounds $1, \dots, |\cands|$, with $r$ denoting an arbitrary round.
Note that tabulation can finish in fewer than $|\cands|$ rounds, if all seats are filled before everyone else has been eliminated or if, at some point, the number of unfilled seats equals the number of remaining candidates.

% ---------------------------------------------------------------------------

\section{Existing Margin Lower Bounding Algorithm}\label{sec:prior-work}

In this paper, we build on the margin lower bound computation algorithm presented by \citet{blom2019toward}.
We refer to this algorithm as \OLD{}, from the initials of the authors and the year it was published.
In this section we provide a high-level overview of \OLD{} and discuss its main
components.
Later, in \autoref{sec:new-algorithm}, we present our new algorithm,
highlighting where it differs from the implementation of \OLD{}.

\subsection{High-Level Overview}\label{sec:PriorOverview}

\begin{algorithm}[t]
\caption{The \OLD{} algorithm for computing a lower bound on the margin of an STV election $\election$ with candidates $\cands$, ballots $\ballots$, number of seats $N$, quota $\quota$, and reported winners $\winners$. We use $\glb$ to denote a running upper limit on the margin lower bound to be returned by the algorithm.}
\label{fig:NewMarginSTV}
\small
\begin{algorithmic}[1]
\Procedure{margin-stv}{$\election = (\cands, \ballots, \seats, \quota, \winners)$}
    \State $F \gets \emptyset$
        \Comment{search frontier, convention: $(l,\, \pi) \in F$, i.e., lower bound and prefix}
    \State $\glb \gets$ \Call{compute-upper-bound}{$\election$}
            \Comment{monotonically non-increasing, see \autoref{sec:UpperBounds}}
    \ForAll{$c \IN \cands \AND a \IN \{0, 1\}$} \Comment{initialise frontier of branch-and-bound search tree}
        \State $\pi \gets [(c, a)]$
        \State $l \gets$ \Call{compute-lower-bound}{$\election,\, \pi,\, \glb$} \Comment{see \autoref{sec:PriorWorkLBHs}}
        \If{$l < \glb$} \APPEND{$(l,\, \pi)$}{$F$}
        \EndIf
    \EndFor
    \State $\llb \gets \min_l F$
        \Comment{monotonically non-decreasing, smallest $l$ attached to a prefix in $F$}
    \While{$F$~\textbf{not}~empty~\textbf{and}~$\llb < \glb$}
        \State $(l,\, \pi) \gets$ pop arg min$_{l}$ $F$
            \Comment{pop the node with the smallest lower bound, $l$}
        \If{$l \geq \glb$} \textbf{continue}
            \Comment{prune node} \label{l:prune}
        \EndIf
        \State $\mathit{Children} \gets$ \Call{expand-and-evaluate}{$\election,\, l,\, \pi,\, \glb$} \Comment{see \autoref{alg:expand-and-evaluate}}
        \ForAll{$(l',\, \pi') \IN \mathit{Children}$}
            \If{$\pi'$ is a leaf node}
                $\glb \gets \min(\glb,\ l')$ \Comment{update running upper limit} \label{l:update}
            \Else{}~\APPEND{$(l',\, \pi')$}{$F$}
            \EndIf
        \EndFor
        \If{$F$ is non-empty} $\llb \gets \min_l F$ \Comment{update running margin lower bound}\label{l:updatelower}
        \EndIf
    \EndWhile
    \If{$F$ is empty} $\llb \gets \glb$
    \EndIf
    \State \Return $\llb$
\EndProcedure
\end{algorithmic}
\end{algorithm}

\begin{algorithm}[t]
\caption{\OLD{}: expansion and evaluation of a prefix $\pi$ in election $\election$ with lower bound $l_{parent}$ 
and running upper limit \glb{} 
to generate the interesting child orders $(\lbchild,\pi') \in \mathit{Children}$. Interesting orders are those for which $\lbchild < \glb$.  In the following, $\mathrm{seated}(\order)$ denotes the set of candidates that have been seated in $\order$, and $\mathrm{remaining}(\order)$ the set of candidates that remain standing after $\order$.}
\label{alg:expand-and-evaluate}
\small
\begin{algorithmic}[1]
\Procedure{expand-and-evaluate}{$\election = (\cands, \ballots, \seats, \quota, \winners),\, l_\textrm{parent},\, \order,\, \glb$}
    \State $\mathit{Children} \gets \emptyset$
    \ForAll{$c \IN \mathrm{remaining}(\order) \AND a \IN \{0,\, 1\}$}
        \Comment{parallelisable}
        \State $\order' \gets \order \plusplus [(c, a)]$ \Comment{create a new prefix}
        \If{$|\mathrm{seated}(\order')| = N$}
            \State \textbf{mark} $\order'$ as a leaf node
        \ElsIf{$N - |\mathrm{seated}(\order')| = |\mathrm{remaining}(\order')|$}
            \State \textbf{mark} $\order'$ as a leaf node
            \State $\mathrm{seated}(\order') \gets \mathrm{seated}(\order') \cup \mathrm{remaining}(\order')$
        \EndIf
        \If{$\mathrm{seated}(\order') = \winners$} \textbf{continue}
            \Comment{skip reported (original) outcomes}
        \EndIf
        \State $\lbeq_{\order'} \gets$ \Call{elim-quota-lb}{$\order'$} \Comment{compute elimination-quota lower bound, see \autoref{sec:PriorWorkLBHs}}
        \State $\lbheur_{\order'} \gets \max(\lbeq_{\order'},\, l_\textrm{parent})$
        \If{$\lbheur_{\order'} \geq \glb$}
            \textbf{continue}
            \Comment{prune node}
        \EndIf
        \State $\lbchild \gets$ \Call{distance-minlp}{$\election,\, \order',\, \lbheur_{\order'},\, \glb$} \Comment{see \autoref{sec:PriorWorkDistanceTo}}
        \If{$\lbchild = \bot$}
            \textbf{continue} \Comment{MINLP was infeasible}
        \ElsIf{$\lbchild = \infty$}
            $\lbchild \gets \lbheur_{\order'}$ \Comment{MINLP timed out}
        \EndIf
        \State \APPEND{$(\lbchild, \order')$}{$\mathit{Children}$}
    \EndFor
    \State \Return $\mathit{Children}$
\EndProcedure
\end{algorithmic}
\end{algorithm}

\OLD{} computes a lower bound on an STV margin by representing the space of
possible outcomes for an STV election as a tree, and searching this tree by
branch-and-bound.
Each node in this tree represents a partial or complete order $\order$,
after a series of eliminations and seatings have taken place.
The leaves of this tree represent complete outcomes in which all seats have
been awarded to candidates.
In contrast to methods for computing IRV margins \citep{blom2016efficient}, the
first level of nodes in the tree represent what occurs in the first round of
tabulation, as opposed to the last round, and each non-leaf node captures a
\textit{prefix} of a complete order, as opposed to a \textit{suffix}.
This difference is a consequence of a useful property of IRV contests that is
lacking in STV contests: the tallies of candidates in any tabulation round of
an IRV contest are completely determined by which candidates have already been
eliminated, and do not depend on the \emph{order} of elimination.  This fact is
not true for STV contests.  Different orders can give different tallies because
transfer values can differ based on the order in which candidates are seated.

\autoref{fig:NewMarginSTV} presents the \OLD{} algorithm. 
\autoref{tab:InitialFrontier} shows the first level of nodes that \OLD{} will
construct for the STV election shown in \autoref{tab:EGSTV1}.
It is simply all the possible first round outcomes.

\begin{table}
\caption{The initial state of the \OLD{} search tree for the example  in \autoref{tab:EGSTV1}. A node is created for each of the ten
    possible first round outcomes. The partial order $\pi_6$ denotes the start
    of the reported outcome.}
\label{tab:InitialFrontier}
\centering
\footnotesize
\begin{tabular}{llllllllll}
\toprule
$\pi_1 =$    &
$\pi_2 =$    &
$\pi_3 =$    &
$\pi_4 =$    &
$\pi_5 =$    &
$\pi_6 =$    &
$\pi_7 =$    &
$\pi_8 =$    &
$\pi_9 =$    &
$\pi_{10} =$ \\
$[(A, 0)]$ &
$[(A, 1)]$ &
$[(B, 0)]$ &
$[(B, 1)]$ &
$[(C, 0)]$ &
$[(C, 1)]$ &
$[(D, 0)]$ &
$[(D, 1)]$ &
$[(E, 0)]$ &
$[(E, 1)]$ \\
\bottomrule
\end{tabular}
\end{table}

\OLD{} tracks its progress via a pair of variables, a `lower limit' and an
`upper limit', that define an interval within which there is a valid lower
bound on the margin.  The final lower bound that is returned always lies within
this interval.  The algorithm updates these limits as it progresses, and uses
them to guide decisions for when to explore a branch and when to prune them
(they are the bounds for the branch-and-bound part of the algorithm).  We refer
to these internal variables as `limits' rather than `bounds', to avoid
confusing them with lower and upper bounds on the  exact margin.

If it were possible to explore every branch, the lower bound that would be
returned would be the smallest value found (for the lower bound) across all
leaf nodes.  The `lower limit' progressively tracks the smallest value seen
across all branches so far.  Since this value is always necessarily a lower
bound on the margin, we also refer to it as the \emph{running lower bound}
(\llb{}).

The `upper limit', which we refer to as the \emph{running upper limit}
(\glb{}), is initially set to an upper bound on the margin (which is
guaranteed to be higher than any valid lower bound on the margin), using
methods described in \autoref{sec:UpperBounds}.  As the algorithm progresses,
the \glb{} is updated to be equal to the smallest value found (for the lower
bound) across all \emph{leaf nodes} that are visited. The rationale for this
is that the final lower bound will be a minimum across branches, and once we
reach a leaf node then the lower bound from that branch cannot get any larger.
Thus, the value obtained is the largest possible value that the algorithm might
return. For the STV election in \autoref{tab:EGSTV1}, the initial \glb{}
(obtained from an upper bound) is 65 votes.

At each node that the algorithm visits, with partial or complete order $\pi$, a
lower bound on a manipulation required to achieve an order that starts with
$\pi$ (or that realises $\pi$, if it is a complete order) is computed. This
is found by solving a relaxation of a MINLP for computing minimal manipulations
(\autoref{sec:PriorWorkDistanceTo}) denoted $\DT$, and/or applying
\textit{lower bounding} heuristics (\autoref{sec:PriorWorkLBHs}). If the lower bound $l$
computed for a complete order $\pi$ is smaller than the current \glb, the \glb{}
is replaced with $l$ (Line~\ref{l:update} of \autoref{fig:NewMarginSTV}).
Nodes whose lower bounds are greater than or equal to the
\glb{} are removed from the tree---we do not explore their descendants (Line~\ref{l:prune}).
Consider the first level of nodes  for the STV election in
\autoref{tab:EGSTV1}. \OLD{} computes lower bounds
that range from 0 to 308 votes for these nodes; see
\autoref{tab:InitialFrontierFirstPrune}.

\begin{table}
\caption{Assignment of lower bounds, $l$, to prefixes in \autoref{tab:InitialFrontier} using \OLD{}'s lower bounding heuristics (\autoref{sec:PriorWorkLBHs}), and $\DT$. Nodes whose $l \geq$ the current \glb{} of 65 votes, or for which the $\DT$ MINLP found could not be manipulated with less than 65 votes, are removed from our tree (shaded grey). }
\label{tab:InitialFrontierFirstPrune}
\centering
\footnotesize
\begin{tabular}{lllllllllll}
\toprule
                               &
\textcolor{gray}{$\pi_1 =$}    &
\textcolor{gray}{$\pi_2 =$}    &
\textcolor{gray}{$\pi_3 =$}    &
\textcolor{gray}{$\pi_4 =$}    &
\textcolor{gray}{$\pi_5 =$}    &
                 $\pi_6 =$     &
\textcolor{gray}{$\pi_7 =$}    &
\textcolor{gray}{$\pi_8 =$}    &
\textcolor{gray}{$\pi_9 =$}    &
\textcolor{gray}{$\pi_{10} =$} \\
                             &
\textcolor{gray}{$[(A, 0)]$} &
\textcolor{gray}{$[(A, 1)]$} &
\textcolor{gray}{$[(B, 0)]$} &
\textcolor{gray}{$[(B, 1)]$} &
\textcolor{gray}{$[(C, 0)]$} &
                 $[(C, 1)]$  &
\textcolor{gray}{$[(D, 0)]$} &
\textcolor{gray}{$[(D, 1)]$} &
\textcolor{gray}{$[(E, 0)]$} &
\textcolor{gray}{$[(E, 1)]$} \\
\midrule
Heuristics                       &
\textcolor{gray}{$l_1    = 125$} &
\textcolor{gray}{$l_2    =  58$} &
\textcolor{gray}{$l_3    =  60$} &
\textcolor{gray}{$l_4    = 188$} &
\textcolor{gray}{$l_5    = 255$} &
                 $l_6    =   0$  &
\textcolor{gray}{$l_7    =   0$} &
\textcolor{gray}{$l_8    = 308$} &
\textcolor{gray}{$l_9    = 308$} &
\textcolor{gray}{$l_{10} =   0$} \\
MINLP                        &
                             &
\textcolor{gray}{infeasible} &
\textcolor{gray}{infeasible} &
                             &
                             &
$l_6 = 0$                    &
\textcolor{gray}{infeasible} &
                             &
                             &
\textcolor{gray}{infeasible} \\
\bottomrule
\end{tabular}
\end{table}

BST-19 then repeatedly: (i)~selects the node with the smallest assigned lower
bound; (ii)~\textit{expands} the node, creating a new node for each of the
possible next decisions that could be made (eliminations and elections) and
computing lower bounds for those nodes using heuristics and $\DT$ MINLP;
and (iii)~adds those nodes to the tree if their lower bounds are smaller than
the current \glb{}. \OLD{} does not store the entire search tree, only its
frontier. \autoref{tab:ExampleExpand} shows the result of expanding node
$\pi_{6}$ in \autoref{tab:InitialFrontier}, with $\pi_6$ replaced with nodes
$\pi_{12}$ and $\pi_{18}$. Node $\pi_{18}$ will be the next node to be
expanded. 
If, upon expansion, the smallest lower bound $\min_l F$ attached to the
nodes on the frontier is greater than the current \llb{}, the \llb{} is increased
to this value (Line~\ref{l:updatelower}).
Once there are no expandable nodes on the frontier, the \llb{} is
returned as the margin lower bound. 
In our running example, \OLD{} finds a
lower bound of 65 votes for the STV election in \autoref{tab:EGSTV1}. As this
is equal to our initial upper bound, we have found an exact margin.

\begin{table}
\caption{Expansion of the prefix $\pi_6$ in \autoref{tab:InitialFrontierFirstPrune}. Nodes whose $l \geq$ than the current \glb{} of 65 votes, or for which the $\DT$ MINLP found could not be manipulated with less than 65 votes, are removed (shaded grey). }
\label{tab:ExampleExpand}
\centering
\footnotesize
\begin{tabular}{lllllllll}
\toprule
                               &
\textcolor{gray}{$\pi_{11} =$} &
                 $\pi_{12} =$  &
\textcolor{gray}{$\pi_{13} =$} &
\textcolor{gray}{$\pi_{14} =$} &
\textcolor{gray}{$\pi_{15} =$} &
\textcolor{gray}{$\pi_{16} =$} &
\textcolor{gray}{$\pi_{17} =$} &
                 $\pi_{18} =$  \\
                            &
\textcolor{gray}{$[(C,1),$} &
                 $[(C,1),$  &
\textcolor{gray}{$[(C,1),$} &
\textcolor{gray}{$[(C,1),$} &
\textcolor{gray}{$[(C,1),$} &
\textcolor{gray}{$[(C,1),$} &
\textcolor{gray}{$[(C,1),$} &
                 $[(C,1),$  \\
                                       &
\textcolor{gray}{$\phantom{[}(A, 0)]$} &
                 $\phantom{[}(A, 1)]$  &
\textcolor{gray}{$\phantom{[}(B, 0)]$} &
\textcolor{gray}{$\phantom{[}(B, 1)]$} &
\textcolor{gray}{$\phantom{[}(D, 0)]$} &
\textcolor{gray}{$\phantom{[}(D, 1)]$} &
\textcolor{gray}{$\phantom{[}(E, 0)]$} &
                 $\phantom{[}(E, 1)]$  \\
\midrule
Heuristics                       &
\textcolor{gray}{$l_{11} =  65$} &
                 $l_{12} =  58$  &
\textcolor{gray}{$l_{13} =   0$} &
\textcolor{gray}{$l_{14} = 188$} &
\textcolor{gray}{$l_{15} =   0$} &
\textcolor{gray}{$l_{16} =   0$} &
\textcolor{gray}{$l_{17} = 115$} &
                 $l_{18} =   0$  \\
MINLP                            &
\textcolor{gray}{$l_{11} =  65$} &
                 $l_{12} =  58$  &
\textcolor{gray}{infeasible}     &
                                 &
\textcolor{gray}{infeasible}     &
\textcolor{gray}{infeasible}     &
                                 &
                 $l_{18} =   0$  \\
\bottomrule
\end{tabular}
\end{table}

\subsection{Margin Upper Bounds}\label{sec:UpperBounds}

\OLD{} used two methods to calculate an upper bound on the STV margin.  The
\emph{winner elimination upper bound} (WEUB) was introduced by
\citet{cary2011estimating} for IRV elections and extended
to STV  by \cite{blom2019toward}. The WEUB considers each elimination in the reported (original) election order. For candidate $c$, eliminated in round $r$, we consider each remaining winner $w$. We use the difference between $w$'s tally in  $r$, and the tally of $c$, to compute how many votes we would need to shift away from $w$ so they would be eliminated in round $r$ instead of $c$. The WEUB is the minimum of all these quantities across original losers. 

In elections where all winners have been elected to a seat prior to any eliminations taking place, the WEUB cannot be computed. In this case, \citet{blom2019toward} defined an alternative. Each $w \in \winners$ that was elected to a seat on the basis of their first preference tally in the reported outcome is considered. One way of altering this outcome is to give a reported loser enough additional first preference votes so that their first preference tally reaches a quota. These votes will be taken away from other candidates. The \emph{SimpleSTV upper bound} is the smallest of these quantities across original losers.

For the STV election in \autoref{tab:EGSTV1}, the WEUB and SimpleSTV upper bound are 65 and 188 votes, respectively. In this case, the WEUB finds the tighter bound.

\subsection{Lower Bounding Heuristics}\label{sec:PriorWorkLBHs}

In \OLD{}, $\DT$ was solved for both partial, and complete, election orders $\order$. In the latter, the result was a lower bound on the manipulation required to realise that \textit{complete} sequence of seatings and eliminations. In the former, the result was a lower bound on the manipulation required to realise a sequence that \textit{starts} with $\order$. 
Although not described in the work of \citet{blom2019toward} for brevity, additional lower bounding heuristics were implemented to, in many cases, determine \textit{tighter} lower bounds than  $\DT$. These heuristics were called the \emph{elimination} and \emph{quota} lower bounding rules.

Given an order $\order$, the \emph{elimination lower bound}, $\lbelim_\order$, is a lower bound on the number of ballots we need to change to ensure that each eliminated candidate in $\order$ has the smallest tally in the round they are eliminated. For a candidate $c \in \cands$, eliminated in round $r$ of $\order$, we compute $c$'s minimum tally at that point, $V^{\min}_{\order,c,r}$. We also compute the maximum possible tally of each other \textit{remaining} candidate $c'$ (i.e., that is \textit{still standing}) at the start of round $r$, according to $\order$. We denote the set of candidates still standing at round $r$ as $\stand_{\order,r}$. For $c$ to be eliminated in  $r$, we need their minimum tally at this point to be \textit{less} than the maximum tally of all  other candidates still standing. Otherwise, we need to take votes away from $c$ to make this so.

\paragraph{Computing the minimum tally of $c$ at round $r$ in $\order$.}
Let $\ballots_{\order,c,r}$ denote the set of ballots that \textit{may} be in $c$'s tally at the start of round $r$, provided the seatings and eliminations in rounds 1 to $r-1$ of $\order$ have taken place.  These are all ballots $b \in \ballots$ for which $c$ is first ranked if we exclude all candidates $\cands \setminus \stand_{\order,r}$. In \OLD{}, the contribution of a ballot $b \in \ballots_{\order,c,r}$ to $c$'s minimum tally at round $r$, $V^{\min}_{\order,c,r}$, was either 0, if a candidate elected in round $r' < r$ in $\pi$ appears before $c$ in the ranking, or 1, otherwise. The reason for assigning a value of 0 to the latter set of ballots is that these ballots may have reduced in value by some amount as a result of one or more surplus transfers. As \OLD{} did not reason about what the value of these ballots could be, they were assigned their minimum value of 0 for the purposes of minimum tally computation. 
\begin{equation}
V^{\min}_{\order,c,r} = \sum_{b \in \ballots_{\order,c,r}} 
\begin{cases}
    0 & \text{a candidate elected in round $r' < r$ in $\pi$ appears before $c$ in $b$} \\
    1 & \text{otherwise}
\end{cases} \label{eq:old:tally_min}
\end{equation}

\paragraph{Computing the maximum tally of a $c'$ at round $r$ in $\pi$.} 
Each ballot $b \in \ballots_{\order,c',r}$ contributes a value of 1 to the maximum tally of candidate $c'$ at round $r$, $V^{\max}_{\order,c',r}$. Some of these ballots may have values below 1 when reaching  $c'$. For  maximum tally computation, \OLD{} assigned them their maximum  value of 1. 
\begin{equation}
V^{\max}_{\order,c',r} = |\ballots_{\order,c',r}| \label{eq:old:tally_max}
\end{equation}

If $c$'s minimum tally is \textit{greater} than the maximum tally of one of the candidates still standing, then they cannot possibly be eliminated in round $r$. Thus, we need to change \textit{at least} the following number of votes:
\begin{equation}
\lbelim_{\order,c} = \max_{c' \in \stand_{\order,r} \setminus \{c\} } \left(\frac{V^{\min}_{\order,c,r}  - V^{\max}_{\order,c',r}}{2}\right)^+
\end{equation}
where $(\cdot)^+$ is the positive part function.\footnote{%
This is defined as $(a)^+ = \max(a, 0)$, which has value $a$ if $a \geqslant 0$ and value 0 if $a < 0$.}
For each $c$ vs $c'$ comparison, the change involves giving some of the votes that would reside with $c$ to $c'$.  

This forms an elimination lower bound with respect to candidate $c$, $\lbelim_{\order,c}$. The overall elimination lower bound for $\order$ is obtained by taking the maximum candidate-based elimination lower bound across all candidates eliminated in $\order$. Let $E_\order \subset \cands$ denote the set of candidates eliminated in order $\order$, then:
\begin{equation}
    \lbelim_{\order} = \max_{c \in E_\order} \lbelim_{\order,c}
\end{equation}

\begin{example}
Consider  $\order_{14} = [(\texttt{C},1),(\texttt{B},1)]$ in \autoref{tab:ExampleExpand} for the STV election of \autoref{tab:EGSTV1}. No candidate in this partial order has been eliminated, and so its elimination lower bound is 0.  In  $\order_{11} = [(\texttt{C},1),(\texttt{A},0)]$, candidate $\texttt{A}$ is eliminated in the second round. To compute $\lbelim_{\order_{11},A}$, we need the maximum possible tally of candidates $\texttt{B}$, $\texttt{D}$, and $\texttt{E}$, and the minimum possible tally of $\texttt{A}$, at the start of the second round. 
\[
V^{\min}_{\pi_{11},\texttt{A},r=2} = 250 \quad
V^{\max}_{\pi_{11},\texttt{B},r=2} = 120 \quad
V^{\max}_{\pi_{11},\texttt{D},r=2} = 400 \quad
V^{\max}_{\pi_{11},\texttt{E},r=2} = 460 \quad
\]
Only $V^{\min}_{\pi_{11},\texttt{A},r=2} - V^{\max}_{\pi_{11},\texttt{B},r=2}$ results in a positive value, and so $\lbelim_{\order_{11}} = \lbelim_{\order_{11},\texttt{A}} = 65$ votes.\\
\label{eg:ExampleELB}
\end{example}

For a partial or complete order $\order$, its \emph{quota lower bound} considers all the candidates that are seated in $\order$. Consider a candidate $c$ that is seated in round $r$ of $\order$. If the maximum tally of $c$ at that point is less than a quota, then $c$ cannot possibly have been seated and we need to give extra votes to $c$ to make it so. \OLD{} uses the same method of computing maximum tallies in both the elimination and quota lower bounding rules. The quota lower bound with respect to candidate $c$ in $\order$ is:
\begin{equation}
    \lbquota_{\order,c} = \left( \quota - V^{\max}_{\order,c,r} \right)^+
\end{equation}
If we denote $W_\order$ as the set of candidates seated in $\order$, the  overall quota lower bound for $\order$ is given by:
\begin{equation}
    \lbquota_{\order} = \max_{c \in W_\order} \lbquota_{\order,c,r}
\end{equation}

\begin{example}
Consider again the order $\order_{14} = [(\texttt{C},1),(\texttt{B},1)]$ in \autoref{tab:ExampleExpand} for the STV election of \autoref{tab:EGSTV1}. Two candidates are elected: $\texttt{C}$ in the first round and $\texttt{B}$ in the second. To compute the quota lower bound for each of these candidates, we compute the their maximum tallies in the round in which they are elected. 
\[
V^{\max}_{\pi_{14},\texttt{C},r=1} = 400 \quad
V^{\max}_{\pi_{14},\texttt{B},r=2} = 120 
\]
Using these values, we compute $\lbquota_{\order_{14},\texttt{C}} = 0$ and $\lbquota_{\order_{14},\texttt{B}} = 188$. The first lower bound is what we would expect, as $\texttt{C}$ is elected to a seat in the first round of the reported outcome. Thus, $\lbquota_{\order_{14}} = 188$ votes.\\
\label{eg:ExampleQLB}
\end{example}

For an order $\order$, we denote its \emph{elimination-quota lower bound} as the maximum  of its quota  and elimination lower bounds. The final lower bound we attach to an order $\order$ is the maximum of its elimination-quota lower bound, and the lower bound found by solving  $\DT$ for $\order$. As a result of the way in which this model has been relaxed, by grouping some sequences of eliminations, the model does not enforce constraints requiring each eliminated candidate to have the smallest tally when eliminated. The elimination-quota lower bounding rules take a more fine grained view, to a certain extent, of the sequence of eliminations and seatings. Consequently, they may derive tighter (higher) lower bounds. 

\begin{example} For the two orders we considered in Examples \autoref{eg:ExampleELB} and \autoref{eg:ExampleQLB}, $\order_{11}$ and $\order_{14}$:
\[
\lbelim_{\order_{14}} = 0 \quad
\lbquota_{\order_{14}} = 188 \quad
\lbelim_{\order_{11}} = 65 \quad
\lbquota_{\order_{11}} = 0
\]
Thus, the elimination-quota lower bound for $\order_{11}$ and $\order_{14}$ is 65 and 188 votes. 
\end{example}

\subsection{MINLP for Finding Cheapest Manipulations}\label{sec:PriorWorkDistanceTo}

\citet{blom2019toward} present a MINLP designed to find a minimal manipulation of a set of ballots, $\ballots$, such that a specific election outcome $\pi$ is realised. Linear approximations of the non-linear constraints were used to form a MILP, $\DT$, that was more tractable to solve. This MILP was designed to capture the variant of STV used to elect senators to the Senate in the Australian Federal Parliament. 

In this paper, we consider a different, and more straightforward, variant of STV, the Weighted Inclusive Gregory method. In \autoref{sec:minlp} and \autoref{app:minlp}, we present the MINLP that we use for minimal manipulation computation. Given advances in non-linear solvers since the work of \citet{blom2019toward}, we do not apply linear approximations and solve the model as a MINLP. 

\subsubsection{Relaxed Orders}\label{sec:RelaxedOrders}

Solving the $\DT$ MILP/MINLP  becomes intractable when dealing with long election orders. The concept of a relaxed order $\order$ was introduced, denoted $\tilde{\order}$, in which some of the sequences of eliminations present in $\order$ were grouped or merged. This technique, although used by \citet{blom2019toward} when defining their $\DT$ MILP, was not described in their paper, and only briefly referred to as \textit{batch elimination} in the supplementary materials. The $\DT$ model involved variables for each possible ranking that could appear on a ballot. By reducing the total number of candidates in the election, by merging some candidates, the number of model variables was considerably reduced.  

 Consider an election order $\order$ $=$ [($\texttt{A},$ 0), ($\texttt{C},$ 1), ($\texttt{B},$ 0), ($\texttt{E},$ 0), ($\texttt{F},$ 0) ($\texttt{D}$, 1)]. This order is relaxed  by grouping candidates $\texttt{B}$ and $\texttt{E}$ into one `super' candidate $\texttt{BE}$, producing $\tilde{\order}$ $=$ [($\texttt{A},$ 0), ($\texttt{C},$ 1), ($\texttt{BE},$ 0), ($\texttt{F},$ 0), ($\texttt{D}$, 1)]. Where ($\texttt{BE},$ 0) appears in the order, it represents candidates $\texttt{B}$ and $\texttt{E}$ being eliminated in some sequence---we just don't care about the order in which those events happen. Formally, we apply candidate merging to  sequences of $n > 3$ candidate eliminations $c_1, \ldots, c_{n-1}, c_n$ by grouping candidates $c_1$ to $c_{n-1}$ into a `super' candidate, leaving $c_n$ out of the merge. When merging eliminated candidates, some constraints in the $\DT$ model, concerned with ensuring those candidates have the lowest tally at the point of their elimination, are removed. Merging entire sequences of eliminated candidates into a single candidate produced a relaxation that was too aggressive, resulting in poor lower bounds on the margin. 

\subsubsection{Equivalence Classes}\label{sec:EquivClasses}

The $\DT$ model used in BST-19 uses the concept of equivalence classes to substantially reduce the number of required variables. The model defines variables for each type of ranking that could appear on a ballot, which we call a \textit{ballot type}. Earlier work by \citet{magrino2011computing} on computing IRV margins recognised that for a given partial or complete election outcome, some ballot types behave in the same way (i.e., they move between the same candidates in each round). For a given order $\pi$, the set of possible ballot types is reduced to a set of \textit{equivalence classes}. Variables used to define the number of ballots of each type that are changed to a different type are then expressed over the smaller set of equivalence classes. We retain the use of equivalence classes in our $\DT$ MINLP.

% ---------------------------------------------------------------------------

\section{Improved \textsc{Margin-STV}}\label{sec:new-algorithm}

By building upon \OLD{} \citep{blom2019toward}, we present a new algorithm specifically designed to compute improved \emph{lower bounds} on the margin of STV elections.
We denote this \textsc{margin-stv}.
The original algorithm is outlined in \autoref{fig:NewMarginSTV}.
The overarching structure of the new algorithm remains unchanged from the work of
\citet{blom2019toward}. The new algorithm incorporates: (i)~tighter
elimination-quota lower bound computation with new transfer path reasoning
(\autoref{sec:revisedlowerbounds}); (ii)~a new
lower bounding heuristic---the displacement lower bound
(\autoref{sec:displacement-lb})---designed to reason about what has to change
\textit{after} the seatings and eliminations in a prefix have occurred; and
(iii)~a new dominance rule designed to reduce the space of partial outcomes
\textsc{margin-stv} has to consider (\autoref{sec:lse}). In addition to the two
methods \OLD{} uses to compute initial upper bounds for an STV margin,
\textsc{margin-stv} includes a third approach---denoted ConcreteSTV
(\autoref{sec:ConcreteSTV})---described in detail by \citet{blom2020did} and
\citet{teague2022annexure}.

Our new algorithm makes the following changes to \autoref{fig:NewMarginSTV}:

\begin{enumerate}
\item In the computation of an initial upper bound in Line 3, we take the minimum of the WEUB (\autoref{sec:UpperBounds}), SimpleSTV (\autoref{sec:UpperBounds}), and ConcreteSTV (\autoref{sec:ConcreteSTV}) upper bounds.

    \item We add the following  between Lines 13 and 14, if using the new order dominance rule  (\autoref{sec:lse}):
\begin{algorithmic}\centering
\If{\Call{dominated}{$\order',\, F$}} \textbf{continue}
\EndIf    
\end{algorithmic}

\item In \autoref{alg:expand-and-evaluate}, we add the following line between
    Lines 11 and 12, if the displacement lower bound is activated
    (\autoref{sec:displacement-lb}):
\begin{algorithmic}\centering
\State $\lbdisp_{\order'} \gets$ \Call{displacement-lb}{$\order'$}
\end{algorithmic}
and we change Line 12 to:
\begin{algorithmic}\centering
\State $\lbheur_{\order'} \gets \max(\lbeq_{\order'},\, \lbdisp_{\order'},\, l_\mathrm{parent})$
\end{algorithmic}
\end{enumerate}

\subsection{ConcreteSTV Upper Bounds}\label{sec:ConcreteSTV}

\citet{blom2020did} and \citet{teague2022annexure} describe a method of
computing upper bounds on STV margins denoted \emph{ConcreteSTV}. This approach
seeks to find actual manipulations of ballots that would result in a changed
outcome when tabulating the manipulated ballot profile. In this way, we can
find and test smaller manipulations than those we know are guaranteed to
achieve different winners. ConcreteSTV additionally considers each seated
candidate, and examines manipulations that rob them of votes.\footnote{%
We used the implementation at
\url{https://github.com/AndrewConway/ConcreteSTV} (accessed 06-Feb-2025).}

\subsection{New, and Improved, Lower Bounding Heuristics} \label{sec:revisedlowerbounds}

When computing lower bounds for a given prefix, $\order$, \OLD{} made conservative assumptions regarding the value of ballots when computing the minimum and maximum tallies of candidates. These minimum and maximum tallies were used to compute an elimination-quota lower bound (\autoref{sec:PriorWorkLBHs}). While each ballot starts with a value of 1, that value is reduced when it is transferred as part of a surplus.
In \OLD{}, however, ballots were instantly assumed to have a zero contribution to minimum tallies if they may have passed through a prior surplus transfer, and a contribution of 1 to maximum tallies.

One of our improvements over \OLD{} stems from new functionality that allows us to calculate transfer values, tallies, and ballot values more accurately when computing minimum and maximum candidate tallies.
This is possible due to our closer analysis of \textit{transfer paths}, which is the series of piles a ballot goes through during tabulation.
In STV tabulation, there is one pile of ballots per candidate, and one pile for exhausted ballots.
The \emph{pile} a ballot is in denotes  which candidate's tally it is counted towards in the tabulation process, or if the ballot is exhausted.
As tabulation proceeds, ballots are moved from pile to pile.
At any step in the tabulation process---which includes which, when, and to whom ballots are transferred---a ballot can only be in one pile; however, the information contained in an imagined prefix $\order$ is not always enough to unambiguously reconstruct a tabulation process. This is because the $\order$ does not prescribe \textit{when} candidates achieve a quota's worth of votes, and that our lower bounding heuristics only know \emph{how many} ballot manipulations are being considered, but not exactly \emph{which} ballots are considered for this manipulation.
Take, for example, the case where $\order = [(\texttt{A}, 1), (\texttt{B}, 1), (\texttt{C}, 1)]$, i.e., we are seating candidates \texttt{A}, \texttt{B} and \texttt{C} in sequence, across rounds 1, 2, and 3.  At the start of round 2, we cannot easily infer whether candidate \texttt{B} reached a quota before \texttt{A} got seated (meaning all ballots transferred from \texttt{A} would skip over \texttt{B}) or whether \texttt{B} reached a quota thanks to ballots transferred from \texttt{A}. It is similarly unclear when \texttt{C} reaches their quota, unless we know which ballots we are manipulating to try and realise $\order$. 

\subsubsection{Transfer Paths.}
There is ambiguity when trying to reconstruct a tabulation process.
The \emph{tail} of a ballot $b$, given a prefix $\order$ and round $r$, is the order of remaining candidates that $b$ \emph{can} (but not necessarily will) be transferred through as the tabulation continues starting with round $r$.
We define it as:\footnote{%
This operation has worst-case time complexity $\bigo(|B| \times |\order|) = \bigo(|\cands|^2)$.
If we calculate this incrementally for each new candidate added to the prefix, we have that the tail function is $\bigo(|\cands|)$ for each node.
}
\begin{equation}
    \tail_{\order,b,r} = \left[x_i \mid 1 \leq i \leq m \textrm{~and~} x_i \not\in \{c_{\order,1}, \dots, c_{\order,{r-1}}\} \right], \quad\mbox{where~} [x_1, \dots, x_m] = b \label{eq:tail}
\end{equation}
where $c_{\order,i}$ denotes the candidate being elected or eliminated in position $i$ of order $\order$. 

\begin{example}
Consider again the order $\order = [(\texttt{A}, 1), (\texttt{B}, 1), (\texttt{C}, 1)]$. For the ballot $ b = [\texttt{A}, \texttt{B}, \texttt{C}]$, and round $r = 2$,  $\tail_{\order,b,r} = [\texttt{B}, \texttt{C}]$. For the ballot $b' = [\texttt{C}, \texttt{A}, \texttt{B}]$, $\tail_{\order,b',r} = [\texttt{C}, \texttt{B}]$.
\end{example}

The pile that ballot $b$ belongs to at the start of round $r$ will be one of the candidates in $\tail_{\order,b,r}$ or the exhausted pile.
The knowledge of what happens in round $r$ (i.e., a candidate is seated or eliminated) gives extra context as to what pile a ballot $b$ could be in.
In particular, the only time piles become ambiguous is when two or more seatings occur in a row in $\order$.

We define $\pile{\order}{b,r}$ as the set of possible piles a ballot $b$ could be in at the start of round $r$ of a prefix $\order$. In the following, $a_{\order,i}$ is 0 when a candidate is eliminated in round $i$ of $\order$ and 1 if a candidate is elected. 
\begin{align}
\pile{\order}{b,r} \ &= \ \begin{cases}
    \{ \mathbf{exhausted} \} &\mbox{if } \tail_{\order,b,r} = \emptyset \\
    \{x_1, \dots, x_m, \mathbf{exhausted}\} &\mbox{if~} a_{\order,r-1} = a_{\order,r} = 1 \mbox{, where~} \tail_{\order,b,r} = [x_1, \dots, x_m]\\
    \{ x_1 \} &\mbox{otherwise, where~} \tail_{\order,b,r} = [x_1, \dots, x_m]
\end{cases} \label{eq:pile}
\end{align}

We can now define what ballots \textit{must} be in a given candidate's pile in a given round, and which ballots \textit{maybe} in their pile. 
\stepcounter{equation}
\begin{align}
\ballots^{\mathrm{must}}_{\order,c,r} \ &= \ \{ b \mid b \in \ballots \textrm{~where~} \{c\} = \pile{\order}{c,r}\} \tag{\theequation a} \label{eq:cand_pile_must} \\
\ballots^{\mathrm{maybe}}_{\order,c,r} \ &= \ \{ b \mid b \in \ballots \textrm{~where~} c \in \pile{\order}{c,r}\} \tag{\theequation b} \label{eq:cand_pile_maybe}
\end{align}

\begin{table}[t]
\centering
\caption{Example of how tail and pile evolves for different ballots $b$ and rounds $r$ of a prefix $\order$. }
\label{tab:tail_and_pile_example}
\begin{tabular}{c|lcccc}
\toprule
        & \multicolumn{5}{c}{prefix $\order = $ [(\texttt{A},0), (\texttt{B},1), (\texttt{C},1), (\texttt{D},0)]} \\
ballot $b$           &    & $r=1$ & $r=2$ & $r=3$ & $r=4$ \\
\midrule
\texttt{[A, D]}    & tail   & \texttt{[A, D]} & \texttt{[D]} & \texttt{[D]} & \texttt{[D]} \\ 
                   & pile   & \{\texttt{A}\} & \{\texttt{D}\} & \{\texttt{D}\} & \{\texttt{D}\} \\
\midrule
% \texttt{[A, B, D]}    & \texttt{[A, B, D]} & \texttt{[B, D]} & \texttt{[D]} & \texttt{[D]}\\
%                       & \{\texttt{A}\} & \{\texttt{B}\} & \{\texttt{D}\} & \{\texttt{D}\} \\
% \midrule
\texttt{[A, C, B]}  & tail  & \texttt{[A, C, B]} & \texttt{[C, B]}  & \texttt{[C]}  & $\emptyset$ \\
                    & pile  & \{\texttt{A}\} & \{\texttt{C}\} & \{\texttt{C}\} & \{\textbf{exhausted}\} \\
\midrule
\texttt{[A, B, C, D]} & tail & \texttt{[A, B, C, D]} & \texttt{[B, C, D]} & \texttt{[C, D]} & \texttt{[D]} \\
                   & pile   & \{\texttt{A}\} & \{\texttt{B}\} & \{\texttt{C, D, }\textbf{exhausted}\} & \{\texttt{D}\} \\
\bottomrule
\end{tabular}
\end{table}

\begin{example}
Let us explore how tail and pile defined for different ballots and different rounds in a prefix $\order$. Consider the prefix $\order =$  [(\texttt{A},0),(\texttt{B},1),(\texttt{C},1),(\texttt{D},0)]. \autoref{tab:tail_and_pile_example} shows how $\tail_{\order,b,r}$ and $\pile{\order}{b,r}$ are computed for different ballots $b$ and rounds $r$ in the prefix. For ballot [\texttt{A}, \texttt{C}, \texttt{B}] and round 2, for example, the tail is [\texttt{C}, \texttt{B}] while the ballot can only be in one pile, that of candidate \texttt{C}. Let us consider what ballots must and maybe in different candidate's piles in different rounds of $\order$. The ballot [\texttt{A}, \texttt{C}, \texttt{B}] must be in candidate \texttt{A}'s pile in round 1, and then in \texttt{C}'s pile in rounds 2 and 3. The ballot [\texttt{A}, \texttt{B}, \texttt{C}, \texttt{D}] must be in candidate \texttt{A}'s pile in round 1, \texttt{B}'s pile in round 2, but then may be in \texttt{C}'s, \texttt{D}'s, or the exhausted pile in round $3$.

Note that our determination of which pile a ballot could be in at a specific round $r$ of a prefix $\order$ \textit{does not} consider events at rounds $r' > r$. In round $3$, we could use information about the remainder of the prefix to infer that any [\texttt{A}, \texttt{B}, \texttt{C}, \texttt{D}] ballot could never be in the exhausted pile. As \texttt{D} is eliminated in the fourth round, we know they could not possibly have had a quota in round $3$, and consequently that ballots of this type will not `skip' over them when \texttt{C}'s surplus is transferred.
\end{example}

\subsubsection{Minimum and Maximum Tallies.}
We have not yet defined \emph{how much} a ballot $b$ contributes to the pile it is in.
As finding the pile of a ballot $b$ at the start of round $r$ of a prefix $\order$ is sometimes ambiguous, the value of a ballot $b$ is similarly sometimes ambiguous.
We denote $\bvaluemax{\order}{b,r}$ and $\bvaluemin{\order}{b,r}$ as the maximum and minimum possible value (between 0 and 1) of ballot $b$ at the start of round $r$ in prefix $\order$. Whenever a candidate $c$ is seated in a round $r$ of a prefix $\order$, there is an associated transfer value $T_{\order,c,r}$. To compute the minimum and maximum value of a ballot $b$ after it has passed through one or more surplus transfers, we need to establish lower and upper bounds on the transfer value associated with each of those transfers. Let $T^{\min}_{\order,c,r}$ and $T^{\max}_{\order,c,r}$ denote a lower and upper bound, respectively, on the transfer value for the seated candidate $c$ in round $r$ of $\order$. We  define these bounds in \autoref{eq:transfervalue_1} and \autoref{eq:transfervalue_2}.
\stepcounter{equation}
\begin{align}
\bvaluemax{\order}{b,1} \ &= \ \bvaluemin{\order}{b,1} \ = \ 1 \tag{\theequation a} \label{eq:bvalue_base} \\
\bvaluemax{\order}{b,r} \ &= \ \begin{cases}
\bvaluemax{\order}{b,r-1} \times T^{\max}_{\order,c,r-1} &\mbox{if~} b \in \ballots^{\mathrm{must}}_{\order,c,r-1} \mbox{~and~} a_{\order,r-1} = 1  \\
\bvaluemax{\order}{b,r-1} &\mbox{otherwise}
\end{cases} \tag{\theequation b} \label{eq:bvalue_max} \\
\bvaluemin{\order}{b,r} \ &= \ \begin{cases}
\bvaluemin{\order}{b,r-1} \times T^{\min}_{\order,c,r-1} &\mbox{if~} b \in \ballots^{\mathrm{maybe}}_{\order,c,r-1} \mbox{~and~} a_{\order,r-1} = 1 \\
\bvaluemin{\order}{b,r-1} &\mbox{otherwise}
\end{cases} \tag{\theequation c} \label{eq:bvalue_min}
\end{align}
When $\order$ contains no seatings, both the minimum and maximum value of a ballot in any round of $\order$ is $1$.

We can now improve upon the equations used by \OLD{} to compute minimum and maximum tallies (\autoref{eq:old:tally_min} and \autoref{eq:old:tally_max}) as follows:
\stepcounter{equation}
\begin{align}
\tallymax{\order}{c,r} \ &= \ \sum_{b \in \ballots^{\mathrm{maybe}}_{\order,c,r}} \bvaluemax{\order}{b,r} \tag{\theequation a}\label{eq:tally_max} \\
\tallymin{\order}{c,r} \ &= \ \sum_{b \in \ballots^{\mathrm{must}}_{\order,c,r}} \bvaluemin{\order}{b,r} \tag{\theequation b}\label{eq:tally_min}
\end{align}

\subsubsection{Bounds on transfer values.}\label{sec:transferbounds}
We use the minimum and maximum tally of a candidate $c$, seated in round $r$ of a prefix $\order$, to establish bounds on their transfer value (\autoref{eq:transfervalue_1} and \autoref{eq:transfervalue_2}).
\stepcounter{equation}
\begin{align}
% T^{\max,\order}_{c,1} \ &= \ T^{\max,\order}_{c,1} \ = \ 1 \tag{\theequation a} \label{eq:transfervalue_1} \\
T^{\max}_{\order,c,r} \ &= \ \frac{\max\left(\quota,\, \tallymax{\order}{c,r}\right) - \quota}{\max\left(\quota,\, \tallymax{\order}{c,r}\right)} \tag{\theequation a} \label{eq:transfervalue_1} \\
T^{\min}_{\order,c,r} \ &= \ \frac{\max\left(\quota,\, \tallymin{\order}{c,r}\right) - \quota}{\max\left(\quota,\, \tallymin{\order}{c,r}\right)} \tag{\theequation b} \label{eq:transfervalue_2}
\end{align}
As we are typically computing lower bounds for prefixes that did not arise in practice, i.e., that do not follow from the cast ballots, candidates may be elected in positions without a quota. We take the max of quota and the actual tally in \autoref{eq:transfervalue_1} and \autoref{eq:transfervalue_2} to arrive at sensible transfer values in these contexts.

\begin{example}
Let us consider our running example from \autoref{tab:EGSTV1} and the prefix $\order = [(\texttt{C},1),(\texttt{E},1),(\texttt{A},0)]$. At the start of the first round, all ballots sit in the pile of their highest ranked candidate, and have a value of 1. As there is no ambiguity around the location and value of ballots, $\ballots^{\mathrm{maybe}}_{\order,c,r=1} = \ballots^{\mathrm{must}}_{\order,c,r=1}$ and $\tallymin{\order}{c,r=1} = \tallymax{\order}{c,r=1}$ for all candidates $c$. Consequently, we can compute an exact transfer value for \texttt{C}, i.e., $T^{\min}_{\order,\texttt{C},1} = T^{\max}_{\order,\texttt{C},1} = 0.396$. At the start of the second round, candidate \texttt{E} will have a minimum tally of $\tallymin{\order}{\texttt{E},r=2} = 350$  votes and a maximum tally of $\tallymax{\order}{\texttt{E},r=2} = 393.56$. The difference arises as the 110 [\texttt{C}, \texttt{E}, \texttt{D}] votes sitting in \texttt{C}'s pile in round 1 may or may not skip over \texttt{D}, when transferred at a value of 0.396 each, depending on when \texttt{D} achieves their quota. We can compute lower and upper bounds on \texttt{E}'s transfer value in round 2 as follows.
\[
T^{\min}_{\order,\texttt{E},2} = \frac{\max(308, 393.56) - 308}
                                      {\max(308, 393.56)} = 0.12,
\qquad\qquad
T^{\max}_{\order,\texttt{E},2} = \frac{\max(308, 350) - 308}
                                      {\max(308, 350)}    = 0.22.
\]
\end{example}

\subsubsection{Revised Elimination and Quota Lower Bounds.}\label{sec:elim-quota-lb}

The elimination-quota lower bound was present in \OLD{}, but has been updated in this paper by replacing the equations used to compute minimum and maximum candidate tallies (\autoref{eq:old:tally_min} and \autoref{eq:old:tally_max}) with new equations (\autoref{eq:tally_min} and \autoref{eq:tally_max}) that reason about transfer paths and transfer values. In \OLD{}, the transfer path concept was not used.
Instead, as soon as a ballot was transferred though a seated candidate it was assumed it was transferred at value 1 when calculating maximum tallies, and 0 when computing minimum tallies. 

\begin{example}
    Let us reconsider the set of prefixes shown in \autoref{tab:ExampleExpand}. The addition of transfer path reasoning results in increased elimination-quota lower bounds for some of these prefixes. For $\order_{15}$, $\order_{16}$, and $\order_{17}$, the new elimination-quota lower bounds increase from 0 to 20, 0 to 150, and 115 to 137, respectively. 
\end{example}

\subsubsection{Displacement Lower Bound.}
\label{sec:displacement-lb}

For a given prefix $\order$, the elimination-quota lower bound considers only the eliminations and seatings present in $\order$. If, by the end of  $\order$, no \textit{new} candidate has been elected, we know that \textit{something} has to change in future rounds. Some original loser will need to be elected in place of an original winner. 
 Consider a prefix $\order$, concluding in round $r-1$, where it is clear that at least one original loser still standing has to displace one of the original winners still 
standing.
In this case, we need to ensure that at least
one of the original losers will not be eliminated before one of the 
original winners.

We compute the displacement lower bound for $\order$, $\lbdisp_\order$, as shown in \autoref{alg:dlb} (\autoref{app:DISP}). First, we check whether $\order$ already changes our reported outcome by  seating a reported loser  or eliminating a reported winner. In both cases, $\lbdisp_\order$ is zero. We then check whether there is scope to change who is elected in subsequent rounds, beyond $\order$. If the number of unfilled seats equals the number of subsequent rounds, all remaining candidates will be automatically seated, and $\lbdisp_\order$ is again zero. We then consider each reported loser $c$ that is still standing (not yet elected or eliminated) at the end of $\order$. We compute three values for this reported loser: the cheapest way we can make sure $c$ is not eliminated before some reported winner still standing ($\mathit{DispCost}_c$); the cheapest way we can ensure $c$ achieves a quota ($\mathit{QuotaCost}_c$); and the cheapest way we can ensure $c$ outlasts enough candidates to be automatically seated in the final round ($\mathit{LeftAtEndCost}_c$). The displacement lower bound with respect to a given reported loser $c$ is:
\begin{equation}
    \lbdisp_{\order,c} = \max\big\{ \mathit{DispCost}_c,\, \min \{ \mathit{QuotaCost}_c,\, \mathit{LeftAtEndCost}_c\} \big\}.
\end{equation}

The displacement lower bound we assign to $\order$, $\lbdisp_\order$, is the smallest of those computed for each reported loser $c$ still standing at the end of $\order$. 

To compute $\mathit{DispCost}_c$, we consider each reported winner $w$ that is still standing at the end of $\order$. We compute the maximum possible tally $c$ could achieve from the end of $\order$ onward, in the context where $w$ is still standing, $\tallymax{\order}{c \prec w,r}$ (\autoref{eq:tallydp}), and contrast this against the minimum tally of $w$ at the end of $\order$, $\tallymin{\order}{w,r}$, computed as per \autoref{eq:tally_min}.  A lower bound on the cost of displacing $w$ with $c$ is equal to half the difference between this maximum and minimum tally. 
\begin{equation}
    \mathit{DispCost}_c \gets \min_{w \in \winners \cap \mathrm{remaining}_\election(\order)} \max \big\{0,\, \frac{1}{2}\big(\tallymin{\order}{w,r} - \tallymax{\order}{c \prec w,r} \big) \big\}
\end{equation}
We define $\tallymax{\order}{c \prec w,r}$, for prefix $\order$, as the maximum total value of all the ballots in which candidate $c$ is ranked before candidate $w$ at the start of round $r$.
\begin{equation}
\tallymax{\order}{c \prec w,r} \ = \ \sum_{b \in \ballots} \begin{cases}
    \bvaluemax{\order}{b,r} &\mbox{if } c \prec w \mbox{ in } \tail_{\order,b,r} \\
    0 &\mbox{otherwise} 
\end{cases} \label{eq:tallydp}
\end{equation}
where $c \prec w$ in a list is true if $c$ appears before $w$ or if only $c$ appears.

For $c$ to be seated, they must either achieve a quota or must never be eliminated. To achieve a quota, their maximum tally $\tallymax{\order}{c,r}$ (\autoref{eq:tally_max}) must reach a quota. We compute the cheapest way for $c$ to achieve a quota ($_c$), and then for $c$ to to be automatically seated in the final round ($\mathit{LeftAtEndCost}_c$). 
\begin{equation}
    \mathit{QuotaCost}_c \gets \max \{ \tallymax{\order}{c,r} - \quota \}
\end{equation}
If there are $N'$ seats left to be filled, and $R$ candidates remaining, $c$ needs to not be eliminated before $L = R-N'-1$ other candidates. We compute and sort the displacement costs between $c$ and each remaining alternate candidate, both reported losers and winners, and take the maximum of the first  $L$ of these displacement costs to form $\mathit{LeftAtEndCost}_c$.

\begin{table}
\caption{Expansion of $\pi_6$ in \autoref{tab:InitialFrontierFirstPrune}, with new lower bounding methods used to compute lower bounds for each node. Nodes whose lower bound is equal to or greater than the current \glb{} of 65 votes, or for which the $\DT$ MINLP found could not be manipulated with less than 65 votes, are removed (shaded grey). }
\label{tab:NewBoundsExample}
\centering
\footnotesize
\begin{tabular}{lllllllll}
\toprule
                               &
\textcolor{gray}{$\pi_{11} =$} &
\textcolor{gray}{ $\pi_{12} =$}  &
\textcolor{gray}{$\pi_{13} =$} &
\textcolor{gray}{$\pi_{14} =$} &
\textcolor{gray}{$\pi_{15} =$} &
\textcolor{gray}{$\pi_{16} =$} &
\textcolor{gray}{$\pi_{17} =$} &
                 $\pi_{18} =$  \\
                            &
\textcolor{gray}{$[(C,1),$} &
\textcolor{gray}{$[(C,1),$}  &
\textcolor{gray}{$[(C,1),$} &
\textcolor{gray}{$[(C,1),$} &
\textcolor{gray}{$[(C,1),$} &
\textcolor{gray}{$[(C,1),$} &
\textcolor{gray}{$[(C,1),$} &
                 $[(C,1),$  \\
                                       &
\textcolor{gray}{$\phantom{[}(A, 0)]$} &
\textcolor{gray}{$\phantom{[}(A, 1)]$}  &
\textcolor{gray}{$\phantom{[}(B, 0)]$} &
\textcolor{gray}{$\phantom{[}(B, 1)]$} &
\textcolor{gray}{$\phantom{[}(D, 0)]$} &
\textcolor{gray}{$\phantom{[}(D, 1)]$} &
\textcolor{gray}{$\phantom{[}(E, 0)]$} &
                 $\phantom{[}(E, 1)]$  \\
\midrule
Heuristics                       &
\textcolor{gray}{$l_{11} =  65$} &
 \textcolor{gray}{$l_{12} =  118$}  &
\textcolor{gray}{$l_{13} =   84$} &
\textcolor{gray}{$l_{14} = 188$} &
\textcolor{gray}{$l_{15} =   65$} &
\textcolor{gray}{$l_{16} =   150$} &
\textcolor{gray}{$l_{17} = 137$} &
                 $l_{18} =   24$  \\
MINLP                            &
\textcolor{gray}{} &
\textcolor{gray}{}   &
\textcolor{gray}{}    &
                                 &
\textcolor{gray}{}     &
\textcolor{gray}{}     &
                                 &
                 $l_{18} =   24$  \\
\bottomrule
\end{tabular}
\end{table}

\begin{example}
    Consider the prefixes in \autoref{tab:ExampleExpand}. We can compute non-zero displacement lower bounds for  $\order_{12}$, $\order_{13}$, $\order_{15}$, and $\order_{18}$, at 118, 84, 65, and 24 votes respectively. The resulting lower bounds computed for the prefixes of \autoref{tab:ExampleExpand} are shown in \autoref{tab:NewBoundsExample}. We are able to prune all children of $\order_6$ except $\order_{18}$. Notably, we are able to avoid solving the $\DT$ MINLP for all but one of the nodes in \autoref{tab:NewBoundsExample}.
\end{example}

\subsection{Leveraging Structural Equivalence}\label{sec:lse}

To improve efficiency, we want to maximise the portion of the alternate-outcome search space \textsc{margin-stv} can ignore. To do so, we introduce an order dominance rule.
 We say that a node $(l, \order)$ is dominated by another $(l'', \order'')$ if $l'' \leq l$ and the relaxed representations of the associated prefixes $\tilde{\order}$ and $\tilde{\order''}$ are the same, $\tilde{\order} \equiv \tilde{\order''}$. When deciding whether to add an $(l, \order)$ to our frontier, $F$,  we check whether $(l, \order)$ is dominated by another node already in $F$, or one that we have expanded before. If so, we do not add it to the frontier.  

This dominance rule relies on comparing the relaxed representations of two orders, and on the following property of our lower bounding heuristics (the displacement and elimination-quota lower bounds): that the contribution of each elimination or election event to the evaluation of the bound is not dependent on the precise order in which candidates have been eliminated or elected prior to the event. The question is, if we have seen an order, $\order''$, with a given relaxed structure, $\tilde{\order''}$, in the past, and we see that structure again in order $\order$, do we need to continue to expand $\order$? If we know the lower bound we attached to the past order  $\order''$, $l''$, is smaller or equal to the lower bound we have attached to $\order$, $l$, then we know that the smallest lower bound we could find for any descendent of $\order''$ will be less than or equal to the smallest lower bound we could find for any descendent of $\order$. The $\DT$ MINLP we create when we add a given sequence of events $\order^*$ to the end of either $\order$ or $\order''$ will be the same. The contribution of each event in the new sequence $\order^*$ to the elimination-quota lower bound for both $\order + \order^*$ and $\order'' + \order^*$ will be the same. The displacement lower bound focuses on what happens in the future of $\order$ and $\order''$, and is independent of the difference that may be present in the precise order in which candidates have been eliminated in these prefixes. Consequently, further exploration of descendants of $\order$ will not result in a complete outcome with a smaller lower bound evaluation than found by exploring descendants of $\order''$.

% ---------------------------------------------------------------------------

\section{$\DT$ MINLP}\label{sec:minlp}

\autoref{app:minlp} presents the full mathematical model of a 
 MINLP designed to find a minimal manipulation to a ballot profile for an STV election such that a specific partial or complete election order is realised. This model assumes the use of Weighted Inclusive Gregory STV. Given an STV election $\election = (\cands, \ballots, \seats, \quota, \winners)$ and a prefix $\order$, the MINLP minimises the number of ballots in $\ballots$ whose rankings are modified in order to realise an election outcome that starts with $\order$. Where $\order$ is a complete outcome, the MINLP minimises the number of ballots we need to modify to realise $\order$. The model is subject to constraints that ensure the total number of ballots remains unchanged by the manipulation, that each eliminated candidate has the smallest tally at the point of their elimination, and that each seated candidate achieves a quota prior to being seated or remains standing at a point where the number of unfilled seats equals the number of remaining candidates. As per \autoref{sec:RelaxedOrders} and \autoref{sec:EquivClasses}, we relax the MINLP by grouping together selected sequences of eliminated candidates into a single batch or `super' candidate, and make use of equivalence classes over ballots to group sets of possible rankings into a smaller set of ballot types. With this relaxation, we remove constraints requiring super candidates to have the smallest tally at the point of their elimination.

% ---------------------------------------------------------------------------

\section{Results}\label{sec:results}

\paragraph{Software.}
We implemented \textsc{margin-stv} in Python 3.8.5.\footnote{%
Our open-source implementation is available at:
\url{https://github.com/michelleblom/pymarginstv}}
All MINLPs were solved using SCIP Optimisation Suite 9.1.1 via the PySCIPOpt 5.1.1 API available as a Python package.
We also used NumPy 1.24.4.
All experiments were run on an Ubuntu 20.04 LTS compute cluster using an Intel Xeon 8260 CPU (24 cores, non-hyperthreaded) with 268.55 GB of RAM.
Each run of the algorithm was allocated 8 processors, 32 GB of memory, and a (wall-clock) time limit of 10,800 seconds (3 hours). When \textsc{margin-stv} expands
a node, the for-loop across lines 3--17 of \autoref{alg:expand-and-evaluate} is parallelised.

MINLP solves terminate if the ceiling of the primal and dual solutions are equal.
For partial prefixes (internal nodes) that do not represent complete outcomes, MINLPs also terminate when the relative gap reaches or falls below 0.01  (i.e., the primal solution is less than 1\% larger than the dual solution) or after 100 seconds.
For complete orders (leaf nodes) that do represent a complete outcome, MINLPs terminate after 150 seconds of solving (no relative gap termination was specified for leaf nodes).
We disabled SCIP's use of relative interior points due to its instability for our problem.

\paragraph{Experiments.}
We compared the performance of \textsc{margin-stv} in terms of runtime and
resulting lower bounds for a suite of real-world STV elections, against \OLD{}.
To evaluate the contribution of the enhancements considered in this
paper---improved elimination-quota lower bounding heuristic, addition of the
displacement lower bounding heuristic, and the new order dominance rule---we
contrasted the performance of \textsc{margin-stv} with all these changes
against variations in which a subset of these enhancements were used.

We evaluated the following methods in this paper:

\begin{description}
    \item[Baseline.] A re-implementation of \OLD{} in Python 3.8.5, with MINLPs solved using SCIP Optimisation Suite 9.1.1 via the PySCIPOpt 5.1.1 API.
    \item[Baseline+U.] A modification of the \textit{Baseline} method with the inclusion of the ConcreteSTV upper bounding method of \autoref{sec:ConcreteSTV}.
    \item[New.] The \textsc{margin-stv} algorithm with transfer path reasoning used in the elimination-quota lower bounding heuristic, but \emph{without} the use of the displacement lower bound or the new order dominance rule.
    \item[New+LSE.] Like \textit{New} but including the new order dominance rule.
    \item[New+DLB.] Like \textit{New} but including the use of the displacement lower bound.
    \item[New+Both.] \textsc{margin-stv} with all enhancements.
\end{description}

\begin{figure}[t]
\centering
\includegraphics[width=\linewidth]{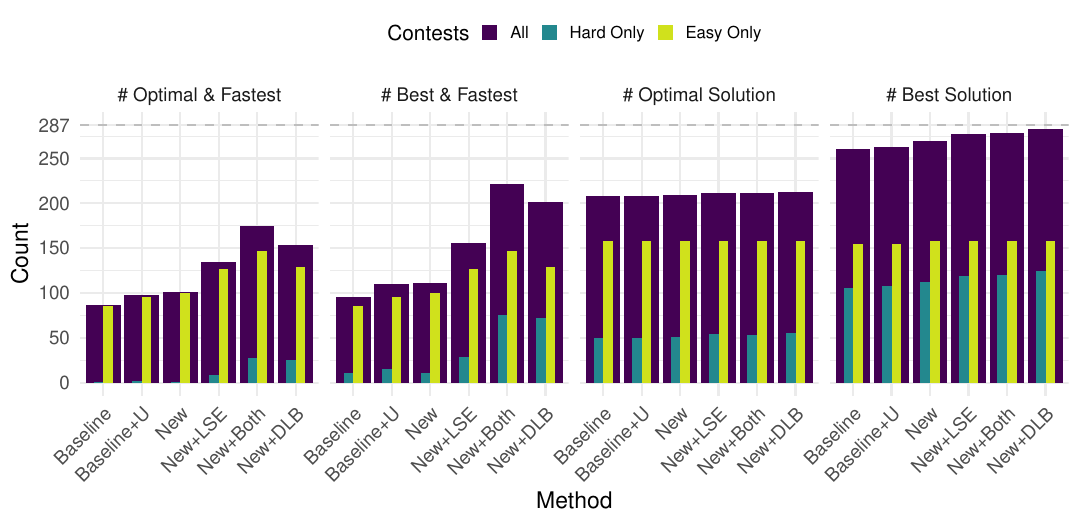}
\caption{Number of contests in each category. `Best Solution' means that the method was one of the methods that obtained the largest margin lower bound (out of all methods on that contest). `Optimal Solution' means that the method returned a margin lower bound that was within 1 ballot of the provided upper bound on the margin. `\& Fastest' means that in addition the method returned a solution in the shortest time (or within 1 second; average of 3 runs).}
\label{plot:summary}
\end{figure}

\paragraph{Data.}
We evaluated the above methods on data from a suite of real-world STV elections
consisting of: 24 contests that were featured in \citet{blom2019toward}; 200
three- and four-seat STV contests held as part of the 2022 local council elections in
Scotland; and 6 two-seat STV contests held as part of the 2016, 2019 and 2022
Australian Senate elections.\footnote{%
Note that we re-imagined the Australian STV contests as using the Weighted
Inclusive Gregory method when in fact they used the Unweighted Inclusive
Gregory method.}
In our analysis, we classified contests as either `hard' or `easy'.
\textit{Easy} contests were those for which the baseline method executes
within 60~seconds and finds a lower bound that is within one ballot of the best
computed upper bound on the margin (the smallest of the WEUB, SimpleSTV, and
ConcreteSTV bounds). All contests not satisfying this property were
classified as \textit{Hard}.

To ensure reliable results, we ran each election contest three times for each method.
We report the mean for the runtime, and the range (if different) for the lower bound found.\footnote{%
For the plots, we used the mean lower bound found as part of the calculations.}
We do not expect vastly different behaviour per run, as there is no inherent
randomness in the algorithm.  The standard error of runtimes across all
contest-method combinations were never larger than 45
seconds, with nearly all (99th percentile) being lower
than 7 seconds.

\subsection{Overall Results}

\autoref{plot:summary} shows, for each method, a count of the number of
election contests where that method, from left to right:
(i)~found the exact margin with the fastest runtime when compared to other methods;
(ii)~found the highest (best) margin lower bound with the fastest runtime when compared to other methods;
(iii)~found the exact margin; and
(iv)~found the highest margin lower bound of those returned by all methods.
Overall, the \textit{New+Both} method appears to more often find the optimal
margin faster in general while \textit{New+DLB} finds the best margin lower bound on the harder contests (where we can't prove optimality); but the differences between the two are not very large.

\begin{figure}[t]
\centering
\includegraphics[width=\linewidth]{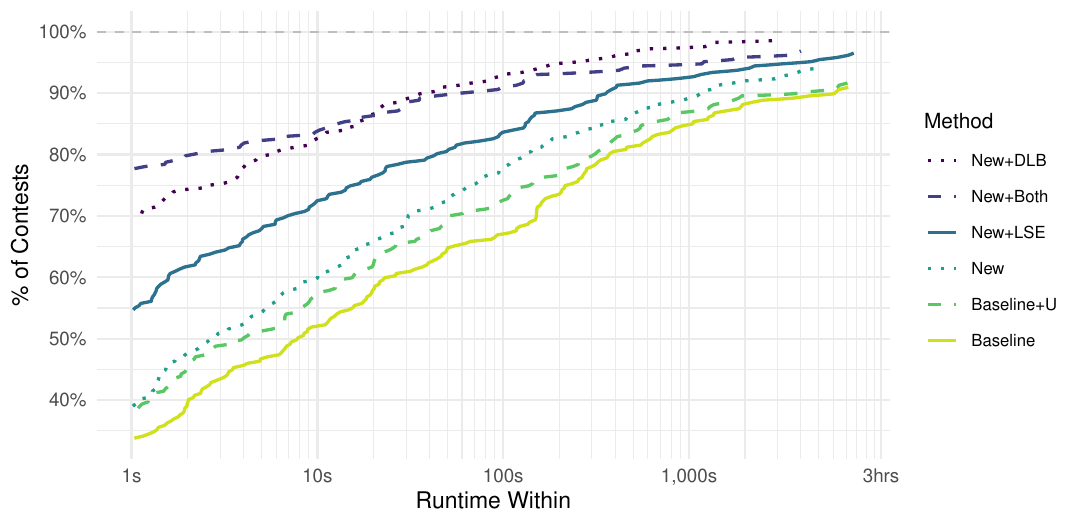}
\caption{For each method, we plot the percentage of election contests $i$ where the runtime of that method is within $x \geq 1$ seconds of the fastest method for that contest, $f_i$, and the resulting lower bound found is no worse than that found by $f_i$.}
\label{plot:within_absolute}
\end{figure}

\autoref{plot:within_absolute} shows the percentage of contests for which the runtime of each method is within $x \geq 1$ seconds of the fastest method (for each contest), across a range of values of $x$. A larger value (of percentage of contests) indicates more computationally efficient performance.
 While \textit{New+Both} is superior on the majority of contests, for the hardest contests \textit{New+DLB} is superior.

\subsection{Selected Contests}

\autoref{tab:datafiles-old-paper} compares the performance of the considered
methods across contests that were featured in \citet{blom2019toward}.  We can
see that for many of these relatively small elections the new method does not
often improve upon the lower bound found, but is usually significantly faster.
The \textit{New+DLB} method has a slight advantage over the \textit{Baseline}
and \textit{New+Both} methods in that it gives slightly better margin lower
bounds in all but one contest. In terms of runtime, \textit{New+Both} is
generally fastest but there are cases where \textit{New+DLB} is
significantly faster. Both new methods are always faster than \textit{Baseline}
except for some very easy contests.

\autoref{tab:improved-instances} records the margin lower bounds found by, and
the mean runtimes of, the \textit{Baseline}, \textit{New+Both}, and
\textit{New+DLB} methods for selected contests from the Scotland 2022 local
council and Australian Senate election datasets. We specifically consider
contests where the difference between the lower bounds found by the new
methods and  \textit{Baseline} was more than one ballot.  Note that on no
contest did the best \textsc{margin-stv} variation (for that contest)
fail to find an equal or higher lower bound than \textit{Baseline}.
Individually, \textit{New+DLB} is usually superior to \textit{New+Both} in
bound but not in runtime.

\begin{table}[t]
\caption{Margin lower bounds found by, and mean runtimes of, the \textit{Baseline}, \textit{New+Both}, and \textit{New+DLB} methods  on contests used in \citet{blom2019toward}; a `---' indicates a timed out. Best results are in bold.}
\label{tab:datafiles-old-paper}
\centering\scriptsize
\begin{tabular}{lrrrr|rrr|rrr}
\toprule
& & & & & \multicolumn{3}{c}{Lower bound found} & \multicolumn{3}{c}{Mean runtime (s)} \\ 
datafile & c & s & q & ub & 
    \rotatebox[origin=r]{-20}{Baseline} &
    \hspace{-1.4cm}\rotatebox[origin=r]{-20}{New+Both} &
    \hspace{-.6cm}\rotatebox[origin=r]{-20}{New+DLB} &
    \rotatebox[origin=r]{-20}{Baseline} &
    \hspace{-5cm}\rotatebox[origin=r]{-20}{New+Both} &
    \hspace{-5cm}\rotatebox[origin=r]{-20}{New+DLB} \\ 
\midrule
      Anderston/C &  9 & 4 & 1381 & 99 &       99 &       99 &  99 & 46.1 & \textbf{26.6} & 49.3 \\ 
      Baillieston & 11 & 4 & 2076 & 105 &      104 &      104 & 104 & 31.1 & \textbf{11.4} & 27.6 \\ 
           Calton & 10 & 3 & 1300 & 376 &      364 &      364 & 364 & 7472.8 & 4083.3 & \textbf{279.6} \\ 
            Canal & 11 & 4 & 1725 & 126 &      125 &      125 & 125 & 59.9 & \textbf{17.8} & 64.1 \\ 
         Craigton & 10 & 4 & 2211 & 75 &       72 &       72 &  72 & 35.2 & \textbf{17.9} & 36.1 \\ 
     Drumchapel/A & 10 & 4 & 1737 & 443 & 359--360 &      \textbf{443} & \textbf{443} & --- & 6131.1 & \textbf{2696.5} \\ 
      East Centre & 13 & 4 & 1816 & 139 &      134 &      134 & 134 & 6631.1 & \textbf{2181.0} & 3524.3 \\ 
     Garscadden/S & 10 & 4 & 2033 & 396 &      396 &      396 & 396 & 8947.0 & 4457.6 & \textbf{2744.4} \\ 
            Govan & 11 & 4 & 1913 & 309 & 277--278 &      \textbf{309} & \textbf{309} & --- & 5063.8 & \textbf{3248.2} \\ 
   Greater Pollok &  9 & 4 & 1737 & 237 &      235 &      235 & 235 & 441.4 & \textbf{73.8} & 90.2 \\ 
         Hillhead & 10 & 4 & 1797 & 105 &      103 &      103 & 103 & 129.0 & \textbf{21.0} & 24.5 \\ 
         Langside &  8 & 3 & 2334 & 233 &      227 &      \textbf{228} & \textbf{228} & 193.0 & \textbf{21.8} & 25.5 \\ 
             Linn & 11 & 4 & 1914 & 218 &      218 &      218 & 218 & 2500.1 & \textbf{1144.2} & 1654.2 \\ 
       Maryhill/K &  8 & 4 & 1981 & 321 &      321 &      321 & 321 & 677.9 & \textbf{114.8} & 273.6 \\ 
       Newlands/A &  9 & 3 & 2164 & 88 &       85 &       85 &  85 & 7.4 & \textbf{4.8} & 5.8 \\ 
       North East & 10 & 4 & 1673 & 421 &      420 &      420 & 420 & 7246.6 & 5225.8 & \textbf{1225.3} \\ 
     Partick West &  9 & 4 & 2549 & 193 &      193 &      193 & 193 & 17.1 & \textbf{8.7} & 11.2 \\ 
    Pollokshields &  9 & 3 & 2392 & 3 &        3 &        3 &   3 & \textbf{1.0} & 1.3 & 1.4 \\ 
      Shettleston & 11 & 4 & 1761 & 353 &      237 & 299--300 & \textbf{318} & --- & --- & --- \\ 
Southside Central &  9 & 4 & 1748 & 229 &      224 &      224 & 224 & 1031.9 & \textbf{249.0} & 277.7 \\ 
       Springburn & 10 & 3 & 1353 & 528 &      400 & 511--512 & \textbf{528} & --- & --- & 4145.5 \\
\midrule
Dublin North & 12 & 4 & 8789 & 211 &      211 &      211 & 211 & 208.9 & \textbf{181.8} & 242.5 \\ 
 Dublin West & 9  & 3 & 7498 & 366 &      366 &      366 & 366 & 33.7 & \textbf{14.5} & 23.5 \\ 
       Meath & 15 & 5 & 10681 & 1113 &    648 & \textbf{     854} & 766 & --- & --- & --- \\ 
\bottomrule
\end{tabular}
\end{table}

\begin{table}[t]
\caption{Margin lower bounds found by, and the mean runtimes of, the \textit{Baseline}, \textit{New+Both}, and \textit{New+DLB} methods for selected election contests from our dataset, where the difference in lower bounds found by the new methods against the baseline was greater than 1 ballot. A `---'  indicates that the method reached the 3 hour timeout. Best results are in bold.}
\label{tab:improved-instances}
\centering\scriptsize
\begin{tabular}{lrrrr|rrr|rrr}
\toprule
& & & & & \multicolumn{3}{c}{Lower bound found} & \multicolumn{3}{c}{Mean runtime (s)} \\ 
datafile & c & s & q & ub & 
    \hspace{-5cm}\rotatebox[origin=r]{-20}{Baseline} &
    \hspace{-5cm}\rotatebox[origin=r]{-20}{New+Both} &
    \hspace{-5cm}\rotatebox[origin=r]{-20}{New+DLB} &
    \hspace{0cm}\rotatebox[origin=r]{-20}{Baseline} &
    \hspace{-5cm}\rotatebox[origin=r]{-20}{New+Both} &
    \hspace{-5cm}\rotatebox[origin=r]{-20}{New+DLB} \\ 
\multicolumn{6}{l}{\quad \textit{Australian Senate}} \\
\midrule
ACT 16 & 22 & 2 & 84923 & 18835 &    42 &              9146 & \textbf{ 9147} & --- & --- & --- \\ 
ACT 19 & 17 & 2 & 90078 & 12939 &   839 &              4186 & \textbf{ 4369} & --- & --- & --- \\ 
ACT 22 & 23 & 2 & 95073 & 11078 &    28 &       \textbf{57} &             19 & --- & --- & --- \\ 
 NT 16 & 19 & 2 & 34010 & 11244 &  2946 &        6835--6836 & \textbf{ 6845} & --- & --- & --- \\ 
 NT 19 & 18 & 2 & 35010 & 15890 &  3033 &        7125--7126 & \textbf{ 7156} & --- & --- & --- \\ 
 NT 22 & 17 & 2 & 34540 & 11412 &   200 & \textbf{655--656} &            178 & --- & --- & --- \\ 
\midrule
\quad \textit{Glasgow 2022} \\
\midrule
Drumchapel/A   & 10 & 4 & 1446 & 327 &      278 & \textbf{323} & \textbf{323} & --- & \textbf{7028.3} & 8243.9 \\ 
East Centre    & 11 & 4 & 1392 & 255 &      241 & \textbf{254} & \textbf{254} & 7267.2 & \textbf{1537.3} & 1946.1 \\ 
Greater Pollok & 11 & 4 & 1774 & 437 & 362--365 &          313 & \textbf{436} & --- & --- & \textbf{4210.3} \\
\midrule
\quad \textit{Other (Scotland 2022)} \\
\midrule
Strathmartine (Dumfries \& G) &  9 & 4 & 1192 & 532 & 428--430 & \textbf{501} & \textbf{501} &    --- &          4048.3 & \textbf{2843.7} \\ 
Torry Ferryhill (Aberdeen)    & 10 & 4 & 1000 & 186 &      164 & \textbf{182} & \textbf{182} & 9100.7 & \textbf{6057.8} &          6674.1 \\ 
\bottomrule
\end{tabular}
\end{table}

\subsection{Discussion}\label{sec:Discussion}

Our results indicate that the new \textsc{margin-stv} algorithm outperforms the
original method (\textit{Baseline}), both in terms of generating tighter margin
lower bounds (in some cases) and in finding those bounds more efficiently (in
almost all cases). The most effective enhancement incorporated into the new
algorithm appears to be the displacement lower bounding heuristic.
\autoref{plot:summary} and \autoref{plot:within_absolute} show marginal
improvement in the quality of lower bounds found and runtime of \textit{New}
over the original method. Recall that \textit{New} represents
\textsc{margin-stv} with only an improved elimination-quota lower bound.
Adding any of the additional improvements (displacement lower bound and the
order dominance rule) on their own substantially improves the performance of
the algorithm. It appears that the benefit of the new order dominance rule is
in general outweighed by the benefit of the displacement lower bound. Both
approaches require some additional computational effort when used. For
contests where the runtime of all methods is substantial (more than a thousand
seconds), it is likely that when both enhancements are used, the effort
expended by the order dominance rule is more often than not simply extending
the runtime without additional benefit.

% ---------------------------------------------------------------------------

\section{Conclusion}\label{sec:conclusion}

In this paper, we present several improvements upon an existing method of
computing lower bounds on the margin of victory for STV elections. Building
upon earlier work on the topic by \citet{blom2019toward}, we introduce new
lower bounding heuristics that, when assessing a lower bound on manipulation
required to realise an outcome that starts in a particular way, provide
\emph{tighter} bounds than earlier methods. This allows us to reduce the size
of the search space of the existing branch-and-bound margin calculation
approach, improving its ability to find better lower bounds within a reasonable
time frame.

This paper presents three specific enhancements over the original method: an
improved elimination-quota lower bounding heuristic; the addition of a
displacement lower bounding heuristic; and a new order dominance rule to reduce
the algorithms search space. We examine the utility of each of these
improvements, finding that the use of the displacement lower bounding heuristic
is responsible for much of the improvement in performance we achieve with the
new algorithm. We show that our new approach is able to find both better lower
bounds than the previous method, and to find these bounds in less time.

One direction for future work is to extend our method to consider a more
nuanced notion of margin of victory to allow for manipulations that change
rankings, add ballots, or remove ballots. Another potential direction is to
enrich our notion of an election prefix or order to include information about
when quotas were achieved by seated candidates. This would remove ambiguity
when computing bounds on transfer values, although it would increase the size
of the space of alternate election outcomes.

% ---------------------------------------------------------------------------

\bibliographystyle{plainnat}
\bibliography{references}

% ---------------------------------------------------------------------------

\FloatBarrier
\clearpage
\appendix

% ---------------------------------------------------------------------------

\section{STV Tabulation Algorithm}\label{sec:STValg}

\autoref{fig:tabulation} outlines the STV tabulation process   under the Weighted Inclusive Gregory method.

\begin{algorithm}[h]
\caption{Pseudocode of the STV tabulation process for an election $\election = (\cands, \ballots, \seats, \quota, \winners)$ under the Weighted Inclusive Gregory method.}
\label{fig:tabulation}
\begin{algorithmic}[1]\footnotesize
\State compute quota $\quota$ (\autoref{eqn:Droop})
\State set tallies according to first-preference votes
\While{seats remain unfilled}
    \If{\# of unfilled seats $=$ \# of remaining candidates} 
        \State seat every remaining candidate
    \ElsIf{no remaining candidate has a tally $\geq \quota$}
        \State eliminate the remaining candidate $e$ with the smallest current tally
        \State transfer each ballot in $e$'s pile to its next ranked remaining candidate
    \Else
        \State seat the candidate $s$ with the current largest tally
        \State calculate $s$'s transfer value $0 \leq \tau < 1$ (\autoref{eqn:TransferValue_simple})
        \State transfer each ballot in $s$'s pile, with a value reduced by $\tau$, to its next ranked remaining candidate with tally $< \quota$
    \EndIf
\EndWhile
\end{algorithmic}
\end{algorithm}

% ---------------------------------------------------------------------------

\section{Displacement Lower Bound: Algorithm}\label{app:DISP}

The algorithm for computing the displacement lower bound for a prefix $\order$ is given in \autoref{alg:dlb}.

\begin{algorithm}[t]
\caption{Displacement lower bound calculation algorithm for an STV election $\election$ and a prefix $\order$ that concludes in round $r-1$, where: $\mathrm{seated}_\election(\order)$ denotes the set of candidates elected to a seat during $\order$; $\mathrm{eliminated}_\election(\order)$  those eliminated during $\order$; $\mathrm{remaining}_\election(\order)$ those still standing after $\order$; and \textsc{sort-take-first}($DPs, L$) a procedure that sorts the list of numbers $DPs$ and returns the first $L$ elements.}
\label{alg:dlb}
\small
\begin{algorithmic}[1]
\Procedure{displacement-lb}{$\election = (\cands, \ballots, \seats, \quota, \winners),\, \order = [(c_1, a_1), \dots, (c_{r-1}, a_{r-1})]$}
\If{$\mathrm{seated}_\election(\order) \not\subset \winners$}
    \Return 0
        \Comment{a reported loser already seated}
\EndIf
\If{$\mathrm{eliminated}_\election(\order) \cap \winners \neq \emptyset$}
    \Return 0
        \Comment{a reported winner already eliminated}
\EndIf
\If{$N - |\mathrm{seated}_\election(\order)| = |\mathrm{remaining}_\election(\order)|$}
    \Return 0
        \Comment{remaining candidates auto-seated}
\EndIf
\State $\mathit{lb} \gets \infty$ \Comment{initialise displacement lower bound}
\State $L \leftarrow |\mathrm{remaining}_\election(\order)| - (N - |\mathrm{seated}_\election(\order)|) - 1$
\ForAll{$c \in \mathrm{remaining}_\election(\order) \setminus \winners$} \Comment{consider all reported losers still standing}
    \State $\mathit{DispCost}_c \gets \min_{w \in \winners \cap \mathrm{remaining}_\election(\order)} \max \big\{0,\, \frac{1}{2}\big(\tallymin{\order}{w,r} - \tallymax{\order}{c \prec w,r} \big) \big\}$
        \Comment{cheapest to displace}
    \State $\mathit{QuotaCost}_c \gets \max \{ \tallymax{\order}{c,r} - \quota \}$
        \Comment{cheapest way to get a quota}
        \Statex $\qquad \quad\triangleright$ cheapest way to never be eliminated (auto-seated)
    \State $DPs \gets [ \frac{1}{2}\big(\tallymin{\order}{c',r} - \tallymax{\order}{c \prec c',r} \big) | \forall c' \in \mathrm{remaining}_\election(\order) \setminus \{c\}]$
    \State $\mathit{LeftAtEndCost}_c \gets \max $\, \textsc{sort-take-first}($DPs, L$)
    \State $\mathit{lb} \gets \min \big\{ \mathit{lb},\, \max\big\{ \mathit{DispCost}_c,\, \min \{ \mathit{QuotaCost}_c,\, \mathit{LeftAtEndCost}_c\} \big\}  \big\}$
\EndFor
\State \Return $\mathit{lb}$
\EndProcedure
\end{algorithmic}
\end{algorithm}

% ---------------------------------------------------------------------------

\section{$\DT$ MINLP}\label{app:minlp}

We present a MINLP designed to find a minimal manipulation to a ballot profile for an STV election such that a specific partial or complete election order is realised. This model assumes the use of Weighted Inclusive Gregory STV. 

\subsection{Indices, Sets, Parameters}
\begin{longtable}{p{25pt}p{410pt}}
$\mathcal{B}$ & Ballots cast in the original election profile. \\[2pt]
$c, \mathcal{C}$ & Candidates. \\[2pt]
$s, \mathbb{S}$ & Ballot types (or signatures). \\[2pt]
$N_s$ & Number of ballots of type $s \in \mathbb{S}$ cast in the original election profile. \\[2pt]
$r, \mathcal{R}$ & Rounds of tabulation. \\[2pt]
$L$ & Last round in which a candidate is either eliminated or elected to a seat with a quota in $\pi$.\\[2pt]
$Q$ & Quota. \\[2pt]
$A_r$ & The subset of candidates still standing at round $r$ of $\pi$ \\[2pt]
$S$ & Number of available seats.\\[2pt]
\end{longtable}

\subsection{Variables}
All non-binary variables are continuous in this model. This is a slight relaxation.

\begin{longtable}{p{25pt}p{410pt}}
$p_s$      & Number of ballots that are modified so that their new type is $s \in \mathbb{S}$. \\[2pt]
$m_s$      & Number of ballots whose original type is $s \in \mathbb{S}$ but have now been changed to a different type. \\[2pt]
$y_s$      & Number of ballots of type $s \in \mathbb{S}$ cast in the new election profile. \\[2pt]
$v_{c,r}$  & Tally of candidate $c$ at the start of round $r$. \\[2pt]
$q_{c,r}$  & Binary variable with value 1 iff the tally of candidate $c$ at the start of round $r$ is at least a quota, and 0 otherwise. \\[2pt]
$nq_{c,r}$ & For convenience, we define a binary $nq_{c,r}$ whose value is 1 iff the tally of candidate $c$ at the start of round $r$ is less than a quota. \\[2pt]
$t_{r}$    & Transfer value applied to ballots leaving an elected candidates' tally in round $r$. These variables are only defined for rounds where a candidate has been seated after achieving a quota, and their ballots distributed at a reduced value. \\
\end{longtable}

\pagebreak
\subsection{Functions}

For each candidate $c$, and round $r$ of $\pi$, we define $f(\pi, c, r)$ as returning a list of tuples ($s$, $v$, $Caveats$) where $s$ denotes a ballot type, $v$ denotes the value of each ballot of that type to $c$, assuming the conditions in $Caveats$ hold, and $Caveats$ a list of binary $q_{c',r'}$ and $nq_{c',r'}$ variables whose values must equal 1 for $c$ to be awarded ballots of type $s$, each with value $v$, in round $r$. If a ballot moves from eliminated candidate to eliminated candidate before it reaches $c$ in $r$, it's value will be 1 ($v = 1$) and $Caveats$ empty. For example, consider the ranking $s$ $=$ ($A$, $B$, $C$) and the order $\pi$ $=$ [($A$, 0), ($D$, 0), ($B$, 0)]. The function $f(\pi, C, 2)$ will return a set of tuples that includes ($s$, 1, []). 

If we know that a ballot will have formed part of one or more surplus transfers before it reaches $c$ in $r$, then its value will equal the product of these transfer values. For example, consider the ranking $s$ $=$ ($A$, $B$, $C$) and the order $\pi$ $=$ [($A$, 1), ($D$, 0), ($B$, 0)], in which $A$'s transfer value was 0.125. The function $f(\pi, C, 2)$ will return a set of tuples that includes ($s$, 0.125, []). For the ranking $s$ $=$ ($A$, $F$, $C$) and order $\pi$ $=$ [($A$, 1), ($D$, 0), ($F$, 1), ($B$, 0)], with $A$ and $F$'s transfer values being 0.125 and 0.05, respectively, the function $f(\pi, C, 3)$ will return a set of tuples that includes ($s$, 0.00625, []).

$Caveats$ will be non-empty in situations where the ballot could have skipped over an elected candidate $c'$ on it's way to $c$, due to $c'$ already having a quota. For the ranking $s$ $=$ ($A$, $F$, $C$) and order $\pi$ $=$ [($A$, 1), ($F$, 1), ($B$, 0)], with $A$ and $F$'s transfer values being 0.125 and 0.05, respectively, the function $f(\pi, C, 2)$ will return a set of tuples that includes both ($s$, 0.00625, [$nq_{F,1}$]) and ($s$, 0.125, [$q_{F,1}$]).

\subsection{Objective}
We minimise the number of ballots modified:
\begin{equation}
    \min \quad \sum_{s} p_s
\end{equation}

\subsection{Constraints}

The number of ballots cast of type $s \in \mathbb{S}$ in the manipulated election profile is equal to the number of ballots originally cast of that type ($N_s$) in addition to the number of ballots of other types modified to have type $s$ ($p_s$), minus the ballots of type $s$ in the original profile changed to a different type ($m_s$). 
\begin{align}
y_s & =  N_s + p_s - m_s & \\
\sum_s p_s & =   \sum_s m_s & 
\end{align}
For candidates $c$ that are elected to a seat in $\pi$ at a round $r' \leq L$:
\begin{align}
v_{c,r} & \geq  Q  q_{c,r} & \forall r < r'  \\
v_{c,r} & \leq  (1 - q_{c,r}) (Q - \epsilon) + |\mathcal{B}| q_{c,r} &  \\
q_{c,r'} & =   1 & 
\end{align}
For rounds $r < L$ in which a candidate $c$ is elected to a seat in $\pi$:
\begin{align}
    t_{r} v_{c,r} & =  v_{c,r} - Q  & \label{cons:tv}
\end{align}
For candidates $c$ that are eliminated in $\pi$ at a round $r \leq L$:
\begin{align}
v_{c,r} & \leq  Q - \epsilon & \\
v_{c,r} & \leq  v_{c',r} & \forall c' \in A_r \setminus \{c\} 
\end{align}
The following constraints define the number of votes in the tally piles of each candidate $c \in \mathcal{C}$ at the start of each round $r$ ($v_{c,r}$) for all rounds $r$ where $c \in \mathcal{D}_r$. 
\begin{align}
    v_{c,0} & =  \sum_s y_s & \forall c \in \mathcal{C} \label{cons:numballots_0}\\
    v_{c,r} & =  v_{c,r-1} + \sum_{(s,v,C) \in f(\pi,c,r-1)} v \,y_s \prod_{x \in C} x & \forall r \in [1, L], c \in A_r \label{cons:value_ballots}
\end{align}

% ---------------------------------------------------------------------------

\section{Additional Results}\label{app:AdditionalResults}

\begin{figure}
\centering
\includegraphics[width=\linewidth]{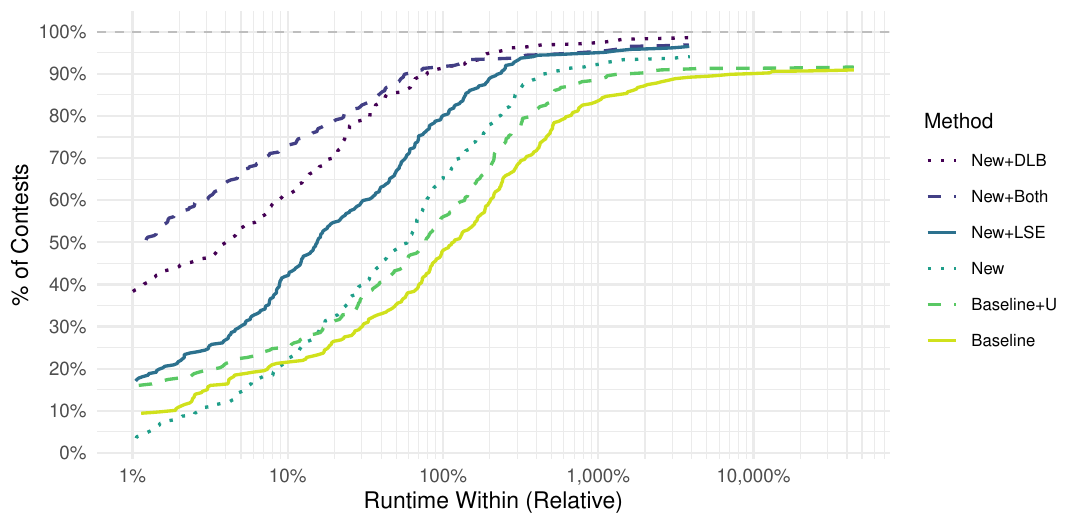}
\caption{For each method, we plot the percentage of election contests $i$ where the runtime of that method is within $x \geq 1$\% of that of the fastest method on that contest, $f_i$, and the resulting lower bound found is no worse than that found by $f_i$.}
\label{plot:within_relative}
\end{figure}

\autoref{plot:within_relative} shows the percentage of contests for which the runtime of each considered method is within $x \geq 1$\% of the fastest method (for each contest), across a range of values of $x$.  A larger value (of percentage of contests) indicates more computationally efficient performance.
We can see that the \textit{New+DLB}  performs the best in these  comparisons.

\end{document}